\documentclass[12pt,psfig,epsfig,twoside]{article}
\usepackage{fleqn,psfig,epsf,espcrc1}
\usepackage{graphicx}

\newcommand{\AmS}{{\protect\the\textfont2
  A\kern-.1667em\lower.5ex\hbox{M}\kern-.125emS}}

\def\shiftdown#1{#1\llap{\lower.04ex\hbox{#1}}}

\begin{document}

\begin{center}

{\bf \large
Field-theoretical description of the multichannel
$\gamma p$ scattering reaction in the $\Delta$ resonance region and
determination of the magnetic moment of the $\Delta^+$ resonance.
}

\end{center}


\vspace{5mm}

\noindent{
{\large \bf A.\ I.\ Machavariani$^a$ $^b$ $^c$ and Amand Faessler $^a$ } 

}

\vspace{5mm}

\noindent{\small

{  \rm $^a$ Institute\ f\"ur\ Theoretische\ Physik\ der\ Univesit\"at\
 T\"ubingen,\newline T\"ubingen\ D-72076, \ Germany}\\

{\rm $^b$ Joint\ Institute\ for\ Nuclear\ Research,\ Dubna,\ Moscow\
region\ 141980,\ Russia}\\

{\rm $^c$ High Energy Physics Institute of Tbilisi State 
University,
University str.  9,  Tbilisi 380086, Georgia  }\\




}

\vspace{0.5cm}


\medskip
\begin{abstract}
{\bf
The cross-sections of the 
$\gamma p-\gamma' N'$, $\gamma p-\pi' N'$ and 
$\gamma p-\gamma'\pi' N'$ reactions are calculated
in the framework of the  field-theoretical  one-particle
 ($\pi,\omega,\rho$-mesons,  nucleon 
and  $\Delta$-resonance) exchange model.   
Unlike  the other relativistic approaches,
our resulting amplitudes of the $\gamma p$ multichannel reactions 
require one-variable covariant vertex functions as input ingredient and 
every diagram  of these amplitudes satisfies the current conservation 
condition in the Coulomb gauge.  
 The complete set of the model independent skeleton
 diagrams for the ${\gamma} p\to\gamma'\pi'N'$ reaction is presented. 
The separable model of the $\pi N$ interaction
is generalized to  construct the  
spin $3/2$ particle propagator of the $\Delta$-resonance.
This procedure allows to obtain the $\pi N-\Delta$ form factor
and   $\Delta$ propagator directly from the  $\pi N$ $P_{33}$ phase shifts.
The numerical calculation of the differential cross section of the 
$\gamma p-\gamma' N'$, $\gamma p-{\pi^o}' N'$ and 
$\gamma p-\gamma'{\pi^o}'N'$ reactions are performed with two different
separable models of the $\Delta$ propagator and with the propagator of 
Breit-Wigner shape. It is demonstrated that the numerical description 
of these reactions in the $\Delta$-resonance region 
are very sensitive to the form of the $\Delta$-propagator. 
The sensitivity of the cross-sections of the $\gamma p\to\gamma'{\pi^o}'p'$ 
 reaction to the magnitude of the $\Delta^+$ magnetic moment is
examined and the
most convenient kinematical region for the determination
of the magnetic moment of the $\Delta^+$-resonance
from the forthcoming data  is indicated. 
}

\end{abstract}


\newpage

\begin{center}
                  {\bf 1. INTRODUCTION}
\end{center}
\medskip

The photon-proton reaction in the  $\Delta$ resonance region of about 
$400MeV$ generates with high probability the 
 following channels: the elastic 
(or proton Compton) scattering, pion photo-production 
($\gamma p\to \pi' N'$), two pion photo-production
($\gamma p\to \pi'\pi' N'$) and  $\gamma p\to \gamma'{\pi}' N'$.
The first two channels, Compton scattering and pi-meson
photoproduction, have a long history.   
Interest to investigate reactions with three-body final
$\gamma \pi p$ states was started with the proposal
to determine the
magnetic moment of the $\Delta^{++}$ resonance in the reaction 
the $\pi^+ p\to \gamma'{\pi^+}' p'$ reaction \cite{Kond}.  
The basic idea of this investigation
is to separate the contribution of the 
$\Delta\to\gamma'\Delta'$ vertex function which in analogy to the
$N-\gamma' N'$ vertex, contains at threshold the magnitude 
of the $\Delta$ magnetic moment. The
contribution of the $\Delta^+\to\gamma'{\Delta^+}'$ vertex function 
in the $\gamma p\to \gamma'{\pi^o}' p'$ reaction was numerically
estimated 
in refs.\cite{M5,Dr1,Dr2} in order to study the dependence of the 
observables on the value of the $\Delta^+$ magnetic moment.
The first data about 
the $\gamma p\to \gamma'{\pi^o}' p'$ reaction
were obtained in a recent experiment by the
A2/TAPS collaboration at MAMI \cite{Kotull} and future experimental
investigations of this reaction are planed by using the Crystal Ball
detector at MAMI \cite{BeckN}.

An other reason to study reactions with $\gamma\pi N$ final
states in the $\Delta$ resonance region is that  
nowadays  a number of different models 
 exist which describe with quite 
good accuracy the experimental data 
of the  $\gamma p\to\pi' N'$ reaction
independent on the $\gamma p\to\gamma' p'$ and $\pi N\to\pi' N'$
channels. 
Therefore, an application of the different two-particle interaction models to
a unified  study of the 
$\gamma N-\gamma \pi N$ and $\pi N-\gamma \pi N$ scattering reactions 
allows us to clarify the dynamical mechanism of the two-body $\pi N$, $\pi \pi$
and $\gamma \pi$, $\gamma N$ interactions.
Moreover,
in  recent calculations  of the $\gamma p$ 
and $\pi N$  reactions
not only the off mass shellness of nucleons and $\Delta$'s
are neglected, but also  the retardation effects are 
omitted, i.e. these calculations were performed in the framework of the 
tree approximation. Nevertheless, after different approximations
and with a different choice of parameters for the tree level vertex functions
authors have reproduced separately the Compton scattering on the
proton, the
pion photoproduction  and the $\gamma p\to\gamma \pi^o N$ reaction 
with satisfactory accuracy. Therefore, the next stage of the
theoretical investigation of the $\gamma p$ reactions
is the unified description of the multichannel $\gamma p$ scattering reactions
 with a minimal number of assumptions and approximations.

The aim of this paper is  the unified investigation of the 
multichannel 
$\gamma p\to\gamma' p'$, $\gamma p\to{\pi^o}' p'$ and 
$\gamma p\to\gamma'{\pi^o}' p'$ reactions 
taking into account the retardation effects and 
investigating the sensitivity of cross sections 
of the $\gamma p\to\gamma'{\pi^o}' p'$ reaction 
to the magnitude
of the $\Delta^+$ magnetic moment $\mu_{\Delta^+}$.
In the  field theoretical formulation considered below
nucleons and $\Delta$'s are defined on mass shell, i. e.
we are not forced to use the approximations connected with
the neglect of the off mass shell variables of  nucleons and  
$\Delta$ resonances. 
In order to demonstrate the complications
generated by the off mass shell behaviour
we compare the usual $N-\gamma'N'$ vertex function
with on mass shell nucleons and $g_M$ the magnetic moment of the nucleon

$$<{\bf p'}_N|J_{\mu}(0)|{\bf p}_N>={\overline u}({\bf p'}_N)
\Biggl(\gamma_{\mu}F_1(t)+
i{ {g_M}\over {2m_N}}\sigma_{\mu\nu}(p_N'-p_N)^{\nu}F_2(t)\Biggr) 
u({\bf p}_N)\eqno(1.1)$$

with the corresponding expression with  off mass shell nucleons
\cite{Koch}

$$\Gamma_{\mu}(p'_N, p_N)
=i\sum_{\alpha,\beta=1,2}\Lambda^{\alpha}(p'_N)
\Biggl(\gamma_{\mu}F_{\alpha,\beta}({p'_N}^2,p_N^2,t)+$$

$$i{ {g_M}\over {2m_N}}\sigma_{\mu\nu}(p_N'-p_N)^{\nu}
G_{\alpha,\beta}({p'_N}^2,p_N^2,t)+(p'_N-p_N)_{\mu}
H_{\alpha,\beta}({p'_N}^2,p_N^2,t)\Biggr)
\Lambda^{\alpha}(p_N),\eqno(1.2)$$

where

$$
\Lambda^{\pm}(p)={{\pm p_{\mu}\gamma^{\mu}+W^2}\over{2W}};\ \ \ 
W=p^2.\eqno(1.3)$$

The twelve form factors in eq. (1.2) depend not only 
on the four momentum transfer $t$ as formfactors in (1.1), 
but also on the off mass shell variables  $p_N^2$ and ${p'}_N^2$.

We note that the dependence  
of the cross section $\gamma p\to\gamma'{\pi^o}' p'$ 
on  the magnetic moment of the $\Delta^+$ resonance
$\mu_{\Delta^+}$ is generated by the 
spin-$3/2$ generalization of the $N-\gamma'N'$ vertex
functions (1.1) or (1.2). 
But the $\Delta-\gamma'\Delta'$ vertex functions with off mass shell
nucleon and $\Delta$ are even more complicated as (1.2). 
Therefore
in  most  phenomenological calculations the off mass shellness of
nucleons and $\Delta$'s is omitted from the beginning.  The accuracy
of this approximation is not clear. 
In the present approach only the vertex functions with on mass shell
nucleons and $\Delta$ resonances are required. Thus we do not have to
worry about the accuracy of the on mass shell approximations.

This paper contains seven sections. In  Sect. 2 
the  construction of the amplitude of the $\gamma p\to \gamma'\pi'N'$
reaction in the old fashioned perturbation theory (or in the
spectral decomposition method over the asymptotic (Fock space) states)
is briefly considered
and the complete set of time ordered diagrams is presented.
The main advantage of this formulation is that as input vertex
functions expressions like (1.1) with only on mass shell
nucleons are required. Sect. 3 deals with consideration of the Coulomb gauge
which insures the validity of current conservation condition for every
diagram if the input vertex functions are gauge invariant.
Besides this we consider in this section  the importance of the
retardation effects for the Born approximations. Section 4 is devoted to 
the problem of construction of the on-mass shell $\Delta$ propagator from
the intermediate $\pi N$ interactions in the old perturbation theory. 
Section 5 deals with a generalization of the separable model 
of the resonance $\pi N$ $t$ matrix
for the case of the spin-$3/2$ particle propagators. 
  In Section 6  the numerical results of
our calculations are  given. The conclusions are presented in Sect. 7.
In Appendix A  all of the input vertex functions with the
corresponding parameterization are listed.

\newpage

\begin{center}
                  {\bf 2. General form 
of the amplitude of the $\gamma N\to\gamma' \pi' N'$  reaction  }
\end{center}

\medskip

\par
The standard definition of the $S$ matrix element of the 
$\gamma N\to\gamma' \pi' N'$ reaction in quantum field theory
\cite{BD,IZ} is 

$$S_{\gamma'\pi' N'-\gamma N}=
(2\pi)^4\ i\ \delta^{(4)}(p'_N+p'_{\pi}+k'_{\gamma}-k_{\gamma}-p_N)
\epsilon^{\mu}({\bf k'},\lambda'){\cal T}_{\mu\nu}
\epsilon^{\nu}({\bf k},\lambda)\eqno(2.1)$$
where  $k_{\mu},\epsilon^{\nu}({\bf k},\lambda)$ and
$k_{\mu}',\epsilon^{\nu}({\bf k'},\lambda')$ indicate the 
four momentum and polarization vector of the initial and the emitted photon,
$p_N=(E_{\bf p_N},{\bf p_N})$, $p_N'=(E_{\bf p'_N},{\bf p'_N})$ and 
$p'_{\pi}=(E_{\bf p_{\pi}},{\bf p'_{\pi}})$ denote the on mass shell 
four-momenta of the
nucleons and pion in the initial and final states  and 
${\cal T}_{\mu\nu}$ is the scattering amplitude of the\
$\gamma N\to\gamma'\pi' N'$ reaction 

$${\cal T}_{\mu\nu}=
<out;{\bf p'}_N{\bf p'}_{\pi}{\bf k'}_{\gamma}\mu |
J_{\nu}(0)|{\bf p}_N>=
\sum_{permutation\ \gamma\ \gamma'\ \pi'}\int d^4x\int d^4y 
e^{ik_{\gamma}'x+ip_{\pi}'y}
$$

$$
<{\bf p'}_N|J_{\mu}(x)\theta(x_o-y_o)j_{\pi'}(y)\theta(y_o)
J_{\nu}(0)|{\bf p}_N>+\ \ \ equal\ time\ commutators
\eqno(2.2)$$
with the well-known time-ordered step function
$\theta(x_o)=1\ \ if\ x_o>0\ \ and\ \ \theta(x_o)=0\  \ if\ x_o<0$.  
The expressions of pi-meson and photon source operators are defined
from the equations of motion for the $\pi$-meson and photon
field operators

$$\Bigl({\partial^{2}}_{x}+m_{\pi}^2\Bigr)\Phi_{\pi}(x)=j_{\pi}(x);
\ \ \ {\partial^{2}}_{x}A_{\mu}(x)=J_{\mu}(x)\eqno(2.3)$$

Substituting the completeness condition $\sum_n |n;in><in; n|={\bf 1}$
in the Eq. (2.2), we get
after integration over $x$ and $y$

$${\cal T}_{\mu\nu}=(2\pi)^6
\sum_{permutation\ \gamma\ \gamma'\ \pi'}\biggl\{\sum_{n,m}
<{\bf p'}_N|J_{\mu}(0)|m;in>
{ {\delta({\bf k'_{\gamma}+p'_N-P_m})}\over
{E_{k'_{\gamma}}+E_{p'_N}-P_m^o+io}}$$
$$<in; m|j_{\pi'}(0)|n;in>
{ {\delta({\bf k'_{\gamma}+p'_N
+p'_{\pi}-P_n})}\over{E_{k'_{\gamma}}+E_{p'_N}+E_{p'_{\pi}}-P^o_n+io}}
<in;n|J_{\nu}(0)|{\bf p}_N>\biggr\}$$
$$+\ \ \ equal\ time\ commutators\eqno(2.4)$$

where $P_n=({P^o}_n,{\bf P}_n)$ denotes the sum of the intermediate
on mass shell particles for the total four-momentum 
$P_n=({P^o}_n,{\bf P}_n)=\sum_i^n p_i $. The indices $\mu,\nu$
in eq. (2.1), (2.4) and every where below 
describe the four-vector operators of photon and indices $\pi$  
denote the isospin quantum number of pion.

Comparing the identical representation of the scattering
amplitude ${\cal T}_{\mu\nu}$ (2.2) and (2.4), we see that

I.\ \ \ \ \ \ The time-ordering procedure in (2.2) is replaced in (2.4) 
by a set of linear propagators which are depend
on external and internal particle energies.

II.\ \ \ \  In expression (2.4) only the sum of all  
three-momenta of the intermediate particles is conserved.

III.\ \ \ Equation (2.4) with one-particle intermediate states  
 contains only  $<{\bf p'}_N|J_{\mu}(0){\bf p"}_N>$,
$<{\bf p'}_N|j_{\pi'}(0){\bf p"}_N>$ etc.
Therefore by construction of the effective, one-particle exchange potential 
based on the three dimensional relativistic 
equations (2.4) only one variable vertex 
functions are required. Thus unlike in the other field-theoretical
formulations, by the calculations  based on eq. (2.4) it is not necessary
to use
some additional approximations in order to obtain one-variable phenomenological
vertex functions.

IV.\ \ \ \ The form of the expression (2.4) is not depending on the
choice of the interaction Lagrangian. Therefore all effective Lagrangian's 
can be incorporated in the present formulation.

V. \ \ \ \ \ In eq. (2.4)  nucleons in the initial ($N$) and in the final
($N'$) states are defined on mass shell.

Expression (2.4) has the form of the spectral decomposition of the 
 amplitude of the $\gamma N\to\gamma' \pi' N'$  reaction by the
complete set of asymptotically free $|n;in>$ states. An analogue spectral
decomposition of the two-body scattering amplitudes (the so called Low
equations)  was investigated in the ref. \cite{Ban,M1,M2} and
the exact linearization procedure of these nonlinear, three-dimensional
equations, was given in ref. \cite{M1,M2}.
\vspace{7mm}

\begin{figure}[htb]
\centerline{\epsfysize=145mm\epsfbox{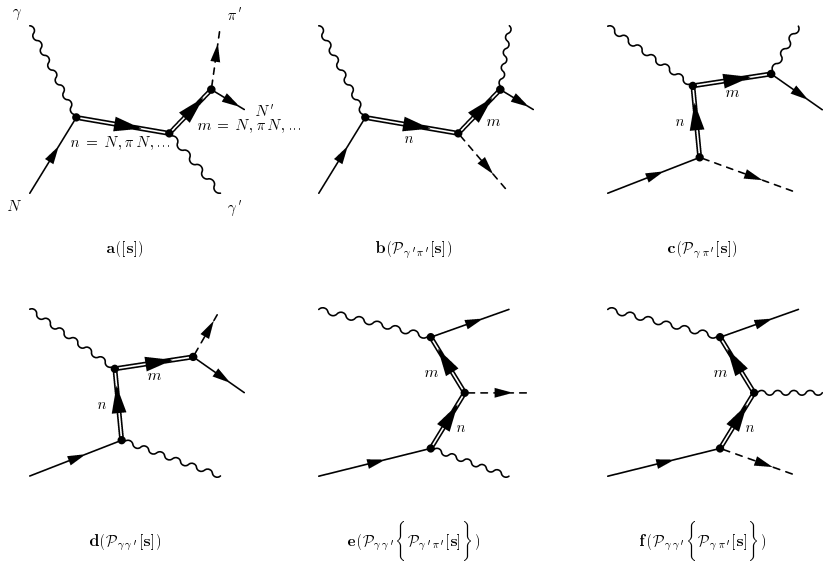}}
\vspace{-8.0cm}
\caption {\footnotesize {\it Diagrammatic representation of the skeleton 
         diagrams of the
         scattering amplitude for the 
         $\gamma N\to\gamma'\pi'N'$
         reaction. These diagrams correspond to all possible permutations
         of the $\gamma,\gamma'$ and $\pi'$ in expression (2.4). The states
         $m$ and $n$ indicate states with a nucleon and a nucleon with an
         arbitrary number of mesons $N,\pi N,...$.}}
\label{fig:one}
\end{figure}
\vspace{7mm}
The diagrammatic representation of equation (2.4) without
equal-time commutators is given in Figure~\ref{fig:one}.
There  only on mass propagators with the total four-momentum 
$P_n=({P^o}_n,{\bf P}_n)=\sum_i^n (\sqrt{m_i^2+{\bf p}_i^2},{\bf p}_i)$
are needed in the intermediate states.
Therefore the diagrams in Figure~\ref{fig:one} are not the Feynman diagrams.
In the time ordered diagram $1a$   
 the incident photon is absorbed first on the  nucleon 
 and   the  $n=N,\pi N,...$ states with $n$
on mass shell  particles are produced. 
This intermediate $n$-particle
state emitted 
first the final photon and transforms into an on shell $m$-particle 
intermediate state. At last this 
$m$-particle state
 transforms in the final pion ($\pi'$) and 
nucleon ($N'$) state. The corresponding chain of the transition matrix  
elements
in expression (2.4) consist of the product of the transition matrices  
$<{\bf p'}_N|J_{\mu}(0)|m;in>$, $<in; m|j_{\pi'}(0)|n;in>$ and
$<in;n|J_{\nu}(0)|{\bf p}_N>$. These transition amplitudes 
are not dependent on the four momentums of the 
$\gamma$ and $\gamma'$ photons and $\pi'$ meson 
and  the values of these four-momenta
are defined  on energy shell
$k_{\gamma}=P_n-p_N$, $k_{\gamma}'=P_n-P_m$ and
$p_{\pi}'=P_m-p_N'$. But in this region the on mass shell condition of 
these particles is not valid i.e. $k_{\gamma}^2=(P_n-p_N)^2\ne 0$,
${k_{\gamma}'}^2=(P_n-P_m)^2\ne 0$  and ${p_{\pi}'}^2=
(P_m-p_N')^2\ne m_{\pi}^2$. Therefore we can assume, that in 
the transition  amplitudes of the expression (2.4) the $\gamma$, $\gamma'$ and
$\pi'$ are defined off mass shell. The initial and final nucleon in (2.4)
are not extracted from the asymptotical states and therefore they remain
on mass shell.

The diagram $b$ in the Figure~\ref{fig:one} is obtained after 
permutation of the final
photon and the pion, i.e.  first the final pion is emitted  
and afterwards the final photon is radiated. 
This permutation procedure is denoted  by
the permutation operator ${\cal P}_{ab}$ of particles $a$ and $b$
in eq. (2.4). In
Figure 1c,1d,1e and 1f  all other possible permutations of $\gamma$,
$\gamma'$ and $\pi'$  of Fig. 1a are given.  This basic diagram
is referred as the direct $s$-channel diagram $[s]$.

In the processes depicted in the Figure~\ref{fig:one}, 
the initial nucleon is absorbed first  and after some intermediate
transformation the final nucleon is emitted last. 
The cluster decomposition
\cite{alf,Ban}  allows us to take into account other 
chronological sequences of the nucleon emission and absorption. 
This procedure is based on the separation
of the connected and disconnected parts of the transition amplitudes. 
In particular,the $\gamma+N\to n$ transition amplitude in Eq. (2.2) or (2.4)
$<in; n|J_{\nu}(0)|{\bf p}_N>$ consists of the  two parts:
$$<in; n|J_{\nu}(0)|{\bf p}_N>=<in;{\bf p"}_N|{\bf p}_N> 
<in; n'|J_{\nu}(0)|0>+<in; n',
{\bf p"}_N|J_{\nu}(0)|{\bf p}_N>_C\eqno(2.5)$$

The first term contains a noninteracting nucleon matrix
element  
$<in;{\bf p"}_N|{\bf p}_N>$ and a term for the
independent transition $\gamma\to n'$.
 Therefore the first part of (2.5) is called
disconnected part of the complete amplitude. The second term in (2.5) 
is connected
and is thus marked by the index $C$. If we take into account the
disconnected part of (2.5), then instead (2.2) we get two terms: 
(Fig. 1a) for the connected vertex 
$<in; n|J_{\nu}(0)|{\bf p}_N>_C$ and the new term

$$
\sum_{permutation\ \gamma\ \gamma'\ \pi'}
\biggl\{ \sum_{n'}\sum_m
\int d^4x\int d^4y 
e^{ik_{\gamma}'x+ip_{\pi}'y}
<{\bf p'}_N|J_{\mu}(x)|m;in>\theta(x_o-y_o)$$
$$<in;m|j_{\pi'}(y)|n'{\bf p}_N;in>\theta(y_o)
<in;n'|J_{\nu}(0)|0>\biggr\}\eqno(2.6a)$$

%
\vspace{5mm}

\begin{figure}[htb]
\centerline{\epsfysize=145mm\epsfbox{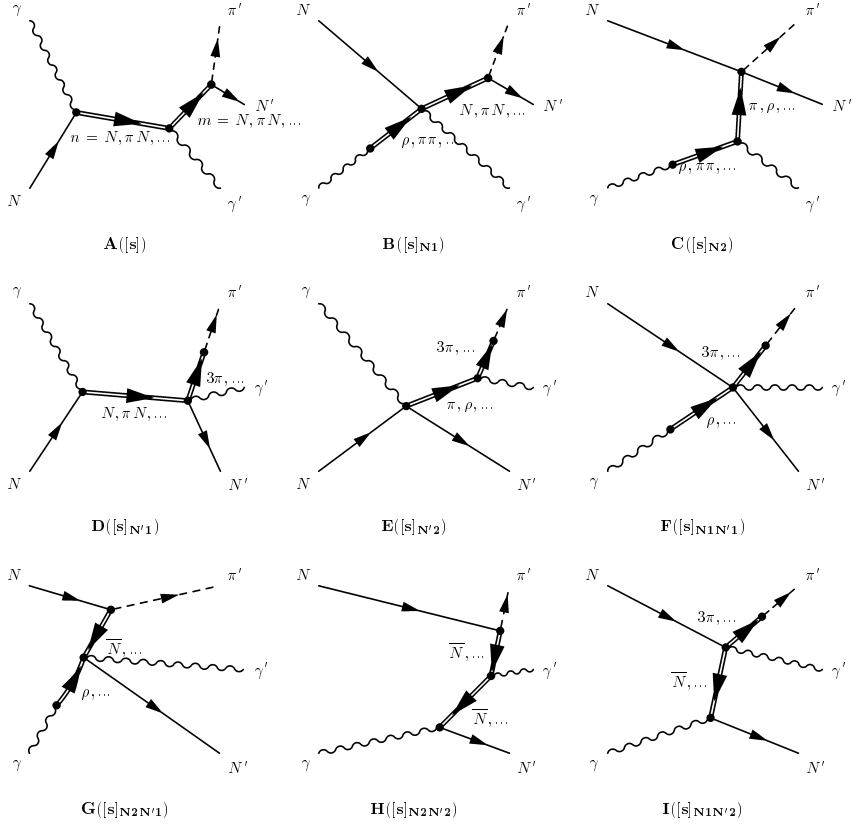}}
\vspace{-4.0cm}
\caption{\footnotesize {\it
         Skeleton diagrams of the
         scattering amplitude for the 
         $\gamma N\to\gamma'\pi' N'$
         reaction with different chronological sequences of the 
         absorption of the initial nucleon (N) and emission of the 
         final nucleon (N'). If one combines the $(\gamma,\pi',\gamma')$
         transposition in Fig.1 and the shifts of the $N$ (diagrams
         B,C), shifts of the $N'$ (diagrams D,E) and the shifts of both $N$
         and $N'$ (diagrams E,F,G,H,I) one obtains all skeleton diagrams
         in the first part of eq. (2.4)
         $<{\bf p'}_N|T\Bigr(J_{\mu}(x)j_{\pi'}(y)J_{\nu}(0)\Bigl)
         |{\bf p}_N>$.}}
\label{fig:two}
\end{figure}
\vspace{2mm}

All possible connected diagrams which are appearing after separation of 
the disconnected parts from the $s$-channel term in the eq. (2.4)
are depicted in the Fig.2. 
These diagrams have 
the fixed time sequence of the
$\gamma$ absorption and the following $\gamma'\pi'$ emission
and all of them are derived from the diagram in the Figure 1a using
the cluster decomposition.
Unlike in the
diagram of  Fig.2A, in the process in the Fig.2B the final photon
is first emitted 
and afterwards the target nucleon $N$ is  absorbed.
If we taken into account the disconnected part of the second matrix element 
$$<in;m|j_{\pi'}(y)|{\bf p}_N n;in>=$$
$$<in;m|j_{\pi'}(y)|{\bf p}_N n';in>_C+
<{\bf p"}_N|{\bf p}_N><in;m'|j_{\pi'}(y)| n';in>,\eqno(2.6b)$$
then we can obtain the process depicted in the Fig.2C, where
the absorption of the target nucleon takes place at the connected,
most remote vertex function. This transposition of the initial nucleon
corresponds to remove $N$ at the second step and is denoted as $[s]_{N2}$.
The diagram in Fig.2B differs from the  diagram in Fig.2A by moving
 $N$  one position the right. It is denoted as $[s]_{N1}$.
In the same manner it is possible to move the final nucleon one or two
positions to the left.
The corresponding diagrams in Fig.2D and Fig2E are marked as 
$[s]_{N'1}$ and $[s]_{N'2}$. 
In the last three diagrams in Fig.2 $N'$  is emitted first and afterwards
the initial nucleon $N$
is absorbed. Therefore in these diagrams an anti-nucleon ${\overline N}$
appears
 in the intermediate states. These diagrams
are so-called $Z$ diagrams and. They describes the time-reversal
processes to diagrams with the 
intermediate nucleon states (Fig.2A,Fig.2B and Fig.2D). If we carry out
the $\gamma,\gamma'$ and $\pi'$ permutations in the same way as it was
performed in the Fig.1 with the $s$-channel diagram, we get $9\times 6=54$ 
diagrams with connected vertex functions. This is a large number of
diagrams, but if we count the number of diagrams with all off-mass shell
particles in the expression $<0|T\biggl(J_{N}(x_1)J_{\gamma}(x_2)
J_{N'}(y_1)J_{\gamma'}(y_2)j_{\pi'}(y_3)\biggr)|>$, then we will get 
even more diagrams $5!=120$. 
 
\medskip

$\underline {\bf Equal-time\ commutators}$ 
in the expression (2.2) have the form

$${\cal Y}\equiv \ equal\ time\ commutators=
\sum_{permutation\ \gamma\ \gamma'\ \pi'}\int d^4x\int dy_o 
e^{ik_{\gamma}'x}$$

$$
<{\bf p'}_N|J_{\mu}(x)\theta(x_o-y_o)\delta(y_o)
\Bigl[ a_{p_{\pi'}}(y_o),
J_{\nu}(0)\Bigr]|{\bf p}_N>\eqno(2.7)$$

where the operator
$$a_{p_{\pi'}}(y_o)=i\int d^3{\bf y}exp({ip_{\pi'}y})
\stackrel{\longleftrightarrow}{\partial_{y_o} } \Phi_{\pi'}(y)\eqno(2.8)$$
transforms into the pion annihilation operator in the asymptotic region
$y_0\to \infty$.

The diagrammatic representation  of the expression (2.7) 
is given in Fig.3, where the  circle relates to the
equal-time commutator ${\cal Y}_{ab}$ between particle $a$ and
particle $b$.

\vspace{5mm}

\begin{figure}[htb]
\centerline{\epsfysize=145mm\epsfbox{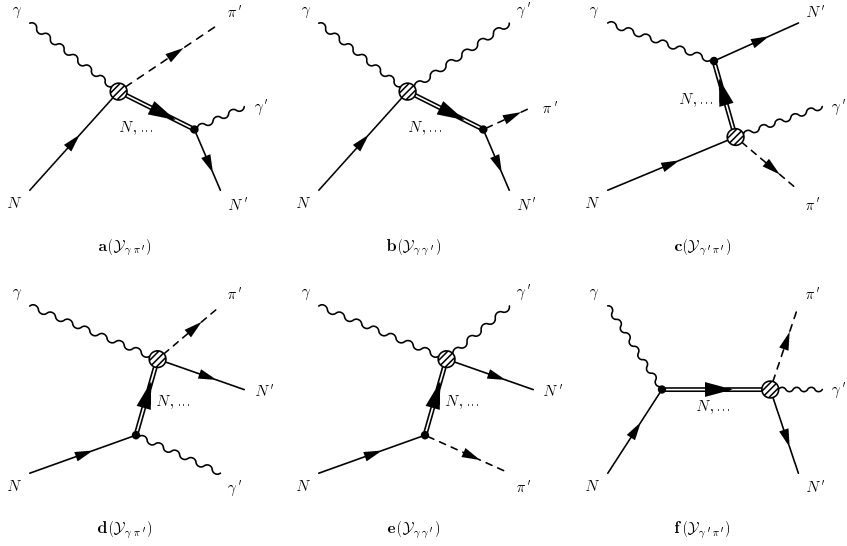}}
\vspace{-8.0cm}
\caption{\footnotesize {\it Diagrammatic representation 
         of the equal-time commutators
         of the   $\gamma N\to\gamma'\pi' N'$ 
         scattering amplitude (2.2), as given by eq. (2.7)}  }
\label{fig:three}
\end{figure}

A corresponding notation of the equal-time
commutators is given in  every diagram in Fig.3.
In the diagrams 3a,3b,3c  the equal-time commutators 
comes first
and afterwards the final nucleon
is emitted.
In  Fig.3c,d,e the target nucleon is absorbed first  and 
after this is the term corresponding to the
equal-time commutator appears.

In order to take into account the connected parts from the transition 
amplitudes in  Fig.3i,e we must carry out the cluster
 decomposition of the expression (2.7). After this procedure every
diagram in Fig.3 produces three additional skeleton diagrams.
These three additional diagrams, which appear after cluster
decomposition of the diagram in Fig.3a,  
are given in  Fig.4.

Thus, we have demonstrated that  the expressions (2.2) or (2.4)
without equal-time terms 
are  expressed by the $9\times 6=54$ diagrams in the Fig.1 and Fig.2
and the equal-time terms (2.7) are depicted 
with the $6\times 4=24$ diagrams  in the Fig.3 and 4. 
The exact form of the
 equal-time commutator is depending on the choice of the 
the  Lagrangian . For example, if we take the interaction Lagrangian with
intermediate vector $V=\rho,\omega$ mesons  
$${\cal L}_{int}=g_{V}/m_{\pi}\epsilon_{\mu\nu\gamma\delta}A^{\mu}(x)
\partial^{\nu}\Phi_{\pi}(x)\partial^{\gamma}V^{\delta}(x),\eqno(2.9) 
$$ 
then the photon source operator is
${j^V}_{\mu}(x)=g_{V}/m_{\pi}\epsilon_{\mu\nu\gamma\delta}
\partial^{\nu}\Phi_{\pi}(x)\partial^{\gamma}V^{\delta}(x)$ and for the
equal-time commutator we get

$$<{\bf p'}_N|\Bigl[{J^V}_{\mu}(x),{a_{\bf p'}}_{\pi}(x_o)\bigr]
\theta(x_o)J_{\nu}(0)|{\bf p}_N>\Longrightarrow 
with\ one\ nucleon\ intermediate\ state\Longrightarrow$$

$$=-i\sum_{{\bf p"}_N}{{g_{V} }\over {m_{\pi} }}\epsilon_{\mu\beta\gamma\delta}
{p'_{\pi}}^{\beta}(p'_N-p_N")^{\gamma}
<{\bf p'}_N|V^{\delta}(0)|{\bf p"}_N>
{ {\delta({\bf k'_{\gamma}+p'_N-p"_N})}\over
{E_{k'_{\gamma}}+E_{p'_N}-E_{p"_N}} }
<{\bf p"}_N|J_{\nu}(0)|{\bf p}_N>.\eqno(2.10)$$

\vspace{7mm}

\begin{figure}[htb]
\centerline{\epsfysize=145mm\epsfbox{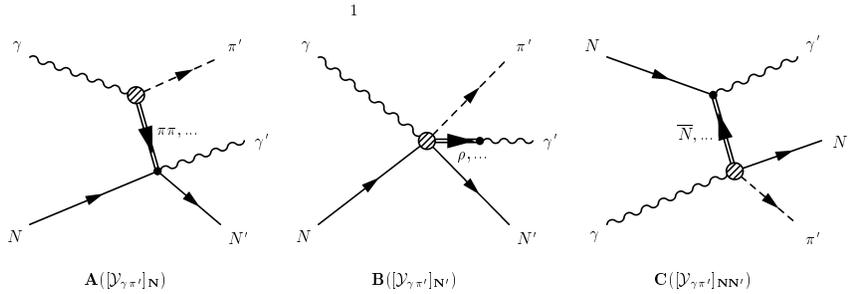}}
\vspace{-12.0cm}
\caption{\footnotesize {\it New type of skeleton diagrams which appear after 
         the cluster decomposition 
         of the basis diagram in the Fig.3a}} 
\label{fig:four}
\end{figure}
\vspace{2mm}

where  
$$<{\bf p'}_N|V^{\delta}(0)|{\bf p"}_N>=
{ {-g^{\delta\sigma}+{p_V}^{\delta}{p_V}^{\sigma}/t}\over{m_V^2-t} }
<{\bf p'}_N|{j^V}_{\sigma}(0)|{\bf p"}_N>\eqno(2.11)$$
and $m_V$ is the $V$-meson mass, ${p_V}^{\mu}={p_N'}^{\mu}-{p_N}^{\mu}$,
$t={p_V}^{\mu}{p_V}_{\mu}$ and the $V$ meson source operator is
$\Bigl({\partial^{2}}_{x}+m_V^2\Bigr)V_{\mu}(x)=j^V_{\mu}(x)$.
Illustration of the equations (2.9) and (2.10) is given in Fig.5.
It is important to note, that the $\gamma\pi\to V=\rho,\omega$ vertex function 
in eq. (2.10) and in the Fig.5 is defined in the tree representation.

\vspace{8mm}

\begin{figure}[htb]
\centerline{\epsfysize=145mm\epsfbox{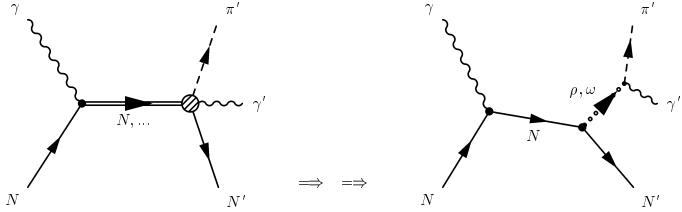}}
\vspace{-12.0cm}
\caption{\footnotesize {\it One $\rho,\omega$ meson exchange diagram 
which is generated
by the equal time commutator (2.10) and calculated according to the effective
Lagrangian (2.9)} }  
\label{fig:five}
\end{figure}
\vspace{2mm}

The $\rho,\omega$-exchange  diagrams are important in the calculation of the
pion photo-production reaction and correspondingly are also important  
 in the $\gamma p\to \gamma'\pi'N'$ reaction. In  refs. \cite{M1,MR} 
it was shown, that the
equal-time commutators generate $t$-channel $\sigma,\rho,\omega$-meson exchange
diagrams also
in the $\pi N$ and $NN$ interactions and these diagrams
 play an essential role
in the pion-nucleon and nucleon-nucleon dynamics. Besides the equal-time 
commutators are very important  in the field-theoretical
formulations including the quark-gluon degrees of freedom.  
Refs.\cite{M1,M2,M3} did show, that in the general theory with quark 
degrees of freedom the form of the scattering equations (2.2) or (2.4)
as well as the form of the diagrams 1,2,3,4 does not change.
All effects of the pure quark-gluon exchange are contained in the 
equal-time commutators.
\medskip

\begin{center}
                  {\bf 3. Coulomb gauge. Gauge invariance
and retardation effect. }
\end{center}

\medskip

In the present formulation 
(which often is also called  old  fashion perturbation theory)
the Coulomb gauge is the natural way to 
exclude the non-physical degrees of freedom of photons and to
insure gauge invariance,
because the three-momentum in  every vertex function of the 
general expression (2.4) is conserved. 
In order to use the Coulomb gauge,
the photon source 
 must be replaced by the transversal source operator
 beginning with the equation of motion and 
 the $S$-matrix reduction formulas\cite{BD} 

$$J_{\mu}(x)\Longrightarrow J_{\mu=\ i\ =1,2,3}^{tr}(x)=J_{i}(x)-
{{\nabla_{i}\partial_{o}}\over {\nabla^2}}J_{o}(x).\eqno(3.1)$$   
 These replacement mean for example, that 
instead of the usual photon-nucleon
vertex function, we get the expression

$$<{\bf p'}_N|J_{i}^{tr}(0)|{\bf p}_N>=
<{\bf p'}_N|J_{i}(0)|{\bf p}_N>-
 {{ \Bigl({\bf p'_N-p_N}\Bigr)_i\Bigl({E_{\bf p'}}_N-{E_{\bf p}}_N\Bigr)}
\over { \Bigl({\bf p'_N-p_N}\Bigr)^2 }}
<{\bf p'}_N|J_{o}(0)|{\bf p}_N>\eqno(3.2)$$

From the current conservation condition $\partial^{{\mu}}J_{\mu}(x)=0$  
follows $\partial^{i}J_{i}^{tr}(x)=0$ \cite{BD}. Thus if we taken into
account that
 in the present
formulation the three-momentum at every vertex function is conserved
${\bf \pm k_{\gamma}=P_m-P_n}$, then we obtain
the current conservation condition
for arbitrary $m,n$ asymptotic states

$${\bf k}_{\gamma}^i <n|J_{i}^{tr}(0)|m>=0\eqno(3.3)$$ 

As a consequence, we see that in the Coulomb gauge  
the validity of the current conservation
of  every term in this formulation is insured.
But the price which we have to pay for this simplification,
is that in Coulomb gauge we have not explicit Lorentz covariance. 
In order to restore the explicit form of the Lorentz-covariance,
one introduces the special form of the polarization 
vector ${\vec{\epsilon} }^{i}({\bf k},\lambda)
\Longrightarrow\epsilon^{\nu}({\bf k},\lambda)$ 
so that the following relation
${{\bf \vec{\epsilon} }}^{i}({\bf k},\lambda)
<n|J_{i}^{tr}(0)|m>=\epsilon^{\nu}({\bf k},\lambda)<n|J_{\nu}(0)|m>$
is valid \cite{BD}. This procedure restores the explicit form of 
Lorentz-invariance of the considered formulation and allows us find
the connections with the Lorentz gauge.

In order to achieve gauge invariance
in this formulation it is not necessary to use 
 additional approximations like the tree approximation with gauge invariant 
combination of terms \cite{Blan}, or the construction of the
approximate  auxiliary gauge-invariance-preserving currents 
\cite{Ohta,Haberz}, or to use the special representation
of the off-mass shell $\Delta$ propagator and the corresponding construction 
 of the  gauge invariant electromagnetic  $\Delta$ vertex
function \cite{Pascalustsa,Castro,Dr2}. For gauge invariance in the old 
Perturbation theory with the Coulomb gauge it is enough to have the gauge 
invariant vertex functions as initial conditions by construction of the 
two-body or the three-body scattering equations.

We emphasize that  gauge invariance in the old perturbation theory was
achieved without tree or special Born approximation
i.e. retardation effects are taken into account  
in the vertex functions and in the set of propagators.
On the other hand in the low and intermediate energy region
the four-momentum transfer $t$ is small. Therefore one can ask the question:
why is it important to take into account the retardation effects
in the Born approximation from the numerical point of view?  

In order to answer this question, let us consider the usual
$\gamma N$ vertex function which is the input vertex in the considered
formulation

$$<{\bf p'}_N|J_{\mu}(0)|{\bf p}_N>={\overline u}({\bf p'}_N)
\Biggl(\gamma_{\mu}F_1(t)+
i{ {g_M}\over {2m_N}}\sigma_{\mu\nu}(p_N'-p_N)^{\nu}F_2(t)\Biggr) 
u({\bf p}_N)\eqno(3.4)$$

In the tree approximation $F_{1,2}(t)= 1$ and
${E_{\bf p'}}_N-{E_{\bf p}}_N\ is\ replaced\ by\ k_{\gamma}$.
For $k_{\gamma}\simeq 350-450MeV$  $t$ is small and 
$F_{1,2}(t)\approx 0.85-0.96$, but 
${E_{\bf p'}}_N-{E_{\bf p}}_N\simeq 40-50MeV.$ Thus the zero component
of the vertex functions (3.4) or (3.2) and the same vertex function in the
tree approximation differ greatly from each other.
This difference is  larger for the Jones-Scadron $\gamma N \Delta$
vertex \cite{Scadron}. Therefore we can conclude, that the gauge invariant
calculations in the tree approximation and gauge invariant calculations
in the Born approximation with retardation effects are not comparable.

\medskip

\begin{center}
                  {\bf 4. On mass shell $\Delta$ extraction from the 
intermediate $\pi N$ interactions.}
\end{center}

\medskip

The extraction of the $\Delta$ resonance from the intermediate
$\pi N$ states may be carried out by replacement of the 
Green function of the interacting $\pi N$ system

$${\cal G}^{\pi N}(E)=
\int d^3{\bf p}{{|\Psi^{\pi N}_{\bf p}><{\widetilde \Psi}^{\pi N}_{\bf p}|}
\over{E-E_{\pi N}({\bf p})-io}}\eqno(4.1)$$
  with the equivalent formula with the intermediate $\Delta$ resonance state 

$${\cal G}^{\pi N}(E)=
\sum_{\Delta}{{|\Psi_{{\bf p}_{\Delta}}><{\hat \Psi}_{{\bf p}_{\Delta}}|}
\over{E-E_{{\bf p}_{\Delta}}-\Sigma_{\Delta}(E)}}
+ \ nonresonant\ part\ \eqno(4.2)$$
as it was done in our previous paper\cite{M5}.

The replacement of the complete Green function (4.1) by the spectral
decomposition formula (4.2) with the intermediate $\Delta$-resonance
state is consistent, because
the representation (4.2)
of the Green function can be considared as definition of the
intermediate on-mass shell $\Delta$ propagator. In addition
the nonresonant contributions in the $P_{33}$ partial waves of 
$\pi N$ amplitudes are small.  
The $\Delta$-propagator in (4.2) is defined  off energy shell because     
the mass operator $\Sigma_{\Delta}(E)$ is depending on the off shell 
parameter $E$. On energy
shell $\Sigma_{\Delta}(E=m_{\Delta}=1232MeV)=const$ we get the 
well known Breit-Wigner shape propagator.
In the $\Delta$ resonance region
the mass operator $\Sigma_{\Delta}(E)$ generates
the $\Delta$ decay width and at $E=m_{\Delta}=1232MeV$
 the following general normalization conditions
for the $\Delta$ propagator \cite{Heller} are valid:


$$
Re\biggl[
E-E_{{\bf p}_{\Delta}}-\Sigma_{\pi N}(E)\biggr]_{E=m_{\Delta}}=0\eqno(4.3a)$$
$$Im\biggl[
E-E_{{\bf p}_{\Delta}}-
\Sigma_{\pi N}(E)\biggr]_{E=m_{\Delta}}=\Gamma_{\Delta}/2\eqno(4.3b)$$
According to the modern   $\pi N$ phase shift analyze \cite{Hoehler}
the Breit-Wigner mass  and width $m_{\Delta}=1232 MeV,\ \ \
\Gamma_{\Delta}=120MeV$ 
differs from the $\Delta$-pole mass and width  
$m_{\Delta}^{pole}=1210 MeV,\ \ \ {\Gamma_{\Delta}}^{pole}=100 MeV$. 
The results of our calculations
 are not sensitive to the above difference of the $\Delta$ mass and width
and in our estimations we will use the magnitudes of 
the Breit-Wigner mass  and width.

In the quantum field theory any arbitrary transition between
the $n+a$ and $m+b$ particle states ($n+a\Longleftarrow m+b$) 
with intermediate $\pi N$ state is described by the formula

$$\sum_{\pi N}<n|j_a(0)|{\bf p}_{\pi}{\bf p}_N>
{{\delta ( {\bf p}_a+{\bf P}_n-{\bf p}_{\pi}-{\bf p}_N ) }\over
{ {E_{\bf p}}_{\pi}+{E_{\bf p}}_N-{P_n}^o-{E_{\bf p}}_{a}+io}}
<{\bf p}_{\pi}{\bf p}_N|j_b(0)|m>$$
$$=\sum_{\pi N}<n|j_a(0)||{\bf p}_{\pi}{\bf p}_N>_{\pi N\ irreducable}
{\cal G}^{\pi N}(E={P_n}^o+{E_{\bf p}}_{a})
<{\bf p}_{\pi}{\bf p}_N|j_b(0)|m>_{\pi N\ irreducable}\eqno(4.4)$$
which according to the replacement (4.1) with the (4.2), can be
rewritten in the form

$$\sum_{\pi N}<n|j_a(0)|{\bf p}_{\pi}{\bf p}_N>
{{\delta ( {\bf p}_a+{\bf P}_n-{\bf p}_{\pi}-{\bf p}_N ) }\over
{ {E_{\bf p}}_{\pi}+{E_{\bf p}}_N-{P_n}^o-{E_{\bf p}}_{a}+io}}
<{\bf p}_{\pi}{\bf p}_N|j_b(0)|m>$$
$$\simeq\sum_{\Delta}\biggl\{<n|j_a(0)\biggr\}_{\pi N\ irreducable}
|\Psi_{{\bf p}_{\Delta}}>
{{\delta ( {\bf p}_a+{\bf P}_n-{\bf P}_{\Delta} )}
\over{E-E_{{\bf p}_{\Delta}}-\Sigma_{\Delta}(E)}}
<{\hat \Psi}_{{\bf p}_{\Delta}}
\biggl\{|j_b(0)|m> \biggr\} _{\pi N\ irreducable}\eqno(4.5)$$
where we have neglected the nonresonant part of $P_{33}$ $\pi N$ 
partial wave contributions.

Formula (4.5) allows to substitute the on mass shell $\Delta$ for the
intermediate $\pi N$ $P_{33}$ partial wave states. Unlike other formulations,
 we have not used
an effective spin $3/2$ Lagrangian 
in order to introduce the intermediate $\Delta 's$.
Any spin $3/2$ Lagrangian has
 free parameters corresponding to  the off-mass shell
degrees of freedom for the massive spin $3/2$ particles.
Therefore in the  approach 
based on the effective spin $3/2$ Lagrangian's, 
additional conditions are necessary in order to determine
the actual off-mass shell behavior of the amplitude.

\medskip

\begin{center}
                  {\bf 5.  Separable model of the $\pi N$ interaction
and propagator of the intermediate $\Delta$ resonance.}

\end{center}

\medskip


\vspace{5mm}

Expression (4.2) allows us to consider the propagator of the intermediate  
$\Delta$ in the following form
$$S^{\alpha\beta}_{{\bf p}_{\Delta}}(E)=
{{ u^{\alpha}({\bf p}_{\Delta}){\overline u}^{\beta}({\bf p}_{\Delta})
}\over{E-E_{{\bf p}_{\Delta}}-
\Sigma_{\pi N}(E)}}
\ {{m_{\Delta}}\over{E_{{\bf p}_{\Delta}} }}\eqno(5.1)$$
where 
$u^{\alpha}({\bf p}_{\Delta})$ is the spinor of the spin $3/2$ particles
with the real(bare) mass $m_{\Delta}$.
In expression (5.1) and everywhere below  we  use the normalization 
condition for fermions from ref.\cite{IZ}. 

In this section our purpose is to determine $\pi N$ scattering
$t$-matrix with the propagator (5.1) in the framework of the 
separable model of the $\pi N$  $P_{33}$
partial waves. The
separable $t$- matrix with intermediate spin $3/2$ propagator
has the following form

$$T({\bf p',p};E)=
g({\bf p'})g({\bf p})(p_{\Delta}-p'_N)_{\alpha}
S^{\alpha\beta}_{{\bf p}_{\Delta}}(E)(p_{\Delta}-p_N)_{\beta}
\eqno(5.2a)$$

where in analogy with  the usual separable model
we have
$$T({\bf p',p};E)
=\lambda g({\bf p'})g({\bf p})
(p_{\Delta}-p'_N)_{\alpha}
{{ u^{\alpha}({\bf p}_{\Delta})
{\overline u}^{\beta}({\bf p}_{\Delta})}\over{
1-\lambda K_{\Delta}(E)}}
(p_{\Delta}-p_N)_{\beta}\eqno(5.2b)$$

and

$$K_{\Delta}(E)=
\int 
{{d^3{\bf q}}\over{(2\pi)^3}}
 {m_N\over{{2E_{\bf q}}_{\pi}{E_{\bf q}}_{N}}}
{{{\bf q}^2   g^2({\bf q}) }\over{E+io-{E_{\bf q}}_{\pi}-
{E_{\bf q}}_{N} }}.\eqno(5.3)$$

According to the separable potential model,
$\lambda$ and $g({\bf p})$ denote the scale and form factor of the 
 $\pi N$ $t$ matrix (5.2b). 





Now in order to find the connection of eq. (5.2a,b) with the ordinary
separable $\pi N$ $t$-matrix for the $P_{33}$ partial wave

$$T_{33}({\bf p',p};E)
=\lambda_{33} {{v({\bf p'})v({\bf p})}\over{1-\lambda_{33} K_{33}(E)}}
\eqno(5.4)$$

$$K_{33}(E)=
\int{{d^3{\bf q}}\over{(2\pi)^3}}
{m_N\over{{2E_{\bf q}}_{\pi}{E_{\bf q}}_{N}}}
{{ v^2({\bf q}) }\over{E+io-{E_{\bf q}}_{\pi}-
{E_{\bf q}}_{N} }},\eqno(5.5)$$

we note, that in the c.m. frame of the $\pi N$ system and on energy shell
(${\bf p}_{cm}^2={{\bf p}_{cm}'}^2\equiv{\bf p}^2$) 
 the following equation is valid
$$(p_{\Delta}-p'_N)_{\alpha}
 u^{\alpha}({\bf p}_{\Delta})
{\overline u}^{\beta}({\bf p}_{\Delta})
(p_{\Delta}-p_N)_{\beta}=
{2\over 3}\biggl({{(p_{\Delta}p_{cm})^2}\over{m_{\Delta}^2}}-p^2_{cm}
\biggr){{\gamma_{\nu}p_{\Delta}^{\nu}+m_{\Delta}}\over {2m_{\Delta}}}$$
$$= {2\over 3}{\bf p}^2_{cm}{{\gamma_{o}+1}\over {2}},\eqno(5.6)$$
where we have used the usual identity
for the $\Delta$ spinor

$$u^{\mu}({\bf p}_{\Delta}){\overline u}^{\nu}({\bf p}_{\Delta})=
-{{p_{\Delta}^{\sigma}
 \gamma_{\sigma}+m_{\Delta}}\over  {2 m_{\Delta}}}
\biggl[g^{\mu\nu}- {1\over 3} \gamma^{\mu}\gamma^{\nu}
-{2\over {3m_{\Delta}^2}}p_{\Delta}^{\mu}p_{\Delta}^{\nu}-
{1\over {3m_{\Delta}}}\bigl(\gamma^{\mu}p_{\Delta}^{\nu}-
\gamma^{\nu}p_{\Delta}^{\nu}\bigr)\biggr].
\eqno(5.7)$$

The projection operator ${{(\gamma_{o}+1)}/ {2}}$ gives $1$ for the 
positive energy fermion states in the rest frame. 
Thus if we define connections between the formfactors

$$g({\bf p})={{v({\bf p})}\over{| {\bf p}|}}\eqno(5.8)$$

we obtain the following relation for on energy shell
$t$ matrices (5.2a) and (5.4)

$${3\over 2}T({\bf p',p};E)
|^{ {\bf p}^2={{\bf p}'}^2}
=T_{33}({\bf p',p};E)
|^{ {\bf p}^2={{\bf p}'}^2}\eqno(5.9a)$$

which   can be continued off energy shell as    

$${3\over 2}T({\bf p',p};E)=T_{33}({\bf p',p};E).\eqno(5.9b)$$

Equations (5.8) and (5.9a,b)  allows us to construct the $\pi N$ $t$ matrix
 (5.2a) or (5.2b) based on the
well known separable model. In particular, 
 form-factors $v({\bf p})$ and  factor $\lambda$
can be defined as solution of the inverse scattering problem or 
can be determined
using the fit of the $\pi N$ phase shifts below $300 MeV$ \cite{GM}.
It is important to note that the $t$-matrix in the separable model (5.4)
is scale
invariant, because the  variation of the $\lambda$ scale parameter 
$\lambda'=\delta\lambda$ can be compensated by the corresponding variation
of the form factors $v'({\bf p})=\delta^{-1/2}v({\bf p})$.  
But for the present calculation based on the formula (4.5) 
the formfactor $g({\bf p})$ is included in the definition of the 
$<n|j_a(0)||\Psi_{{\bf p}_{\Delta}}>$ transition amplitude.
Therefore in (4.5) instead of the complete $\pi N$ $t$-matrix only
the propagator (5.1) is presented  and the scale invariance of the 
separable $\pi N$ $t$-matrix is broken. 

In order to calculate
the  amplitude of the multichannel $\gamma p$ scattering reactions
we  use the normalization condition
of the $\Delta$ propagator (4.3a,b), 
because in the opposite case these amplitudes do not have the correct scale.
The scale invariance of the 
separable $\pi N$ $t$-matrix is not enough to reproduce both conditions
(4.3a,b) for the  $\Delta$ propagator. Therefore in our 
calculation we have used the following models of the $\pi N$
propagator (5.1) and $\pi N-\Delta$ formfactors $g({\bf p})$:

\vspace{5mm}

  A.\ {\bf Scale invariant separable model (MODEL A)\cite{MR}:} 

\vspace{5mm}
 
In this model the $\pi N$ scattering amplitude  reproduce the 
the $P_{33}$ partial wave $\pi N$ phase shifts  up to $300MeV$.
The propagator of the $\Delta$ has the form

$$S^{\alpha\beta}_{{\bf p}_{\Delta}}(E)=
{{ u^{\alpha}({\bf p}_{\Delta}){\overline u}^{\beta}({\bf p}_{\Delta})
}\over{\lambda^{-1}-K_{\Delta}(E)
}}\ {{m_{\Delta}}\over{E_{{\bf p}_{\Delta}} }}\eqno(5.10)$$

where  $K_{\Delta}(E)$ is defined in the eq. (5.3) and 
$$ \lambda^{-1}=Re\Bigl(K_{\Delta}(E=m_{\Delta})\Bigr);\ \ \
g({\bf p})={{\eta}\over{{\bf p}^2+\mu^2}}\eqno(5.11)$$
with  the following choice of the fitting parameters
 $\mu=9m_{\pi},\ \ \ \eta=15.85m_N$. 

The form of the $\lambda^{-1}$ insures the validity
of the condition (4.3a) and the adjustable parameter $\eta$ is fixed 
according to the condition (4.3b).

\vspace{5mm}

 B.\ {\bf Breit-Wigner shape $\Delta$ propagator (MODEL B) \cite{M5}:}

\vspace{5mm}
 
In this model
$E_{{\bf p}_{\Delta}}+\Sigma_{\pi N}(E)\approx
\sqrt{(M_{\Delta}-i{{\Gamma_\Delta}/ 2})^2 +{\bf p}_{\Delta}^2}$
and the $\Delta$ propagator has the following form

$$S^{\alpha\beta}_{{\bf p}_{\Delta}}(E)=
{{ u^{\alpha}({\bf p}_{\Delta}){\overline u}^{\beta}({\bf p}_{\Delta})
}\over{
E-\sqrt{(M_{\Delta}-i{{\Gamma_\Delta}\over 2})^2 +{\bf p}_{\Delta}^2}}}
\ {{m_{\Delta}}\over{E_{{\bf p}_{\Delta}} }}\eqno(5.12)$$

The formfactor $g({\bf p})$ is obtained from the effective $\pi N\Delta$
Lagrangian in tree approximation i. e. it is equal to the $g_{\pi N\Delta}$
coupling constant
$$g({\bf p})=g_{\pi N\Delta}\eqno(5.13)$$
where we have taken the same coupling constant as in ref.\cite{Dr2}
$g_{\pi N\Delta}=1.95/m_{\pi}$.

\newpage

 C.\ {\bf Heller-Kumano-Martinez-Moniz separable potential (MODEL C)
\cite{Heller}:}

\vspace{5mm}

This model was used for the calculation of the $\Delta^{++}$ 
magnetic moment in the $\pi^+p\to\gamma'\pi^+p$ reaction and it reproduces
the $\pi N$ $P_{33}$ phase shifts up to 300 MeV.
In this model the following parameterization is used 
$$E_{{\bf p}_{\Delta}}+\Sigma_{\pi N}(E)\equiv 
M_{\Delta}+{\tilde \Sigma}_{\pi N}(E)
=M_{\Delta}+
{1\over 3}\int {{d^3{\bf q}}\over{(2\pi)^3}}
{{ {\bf q}^2\ h^2({\bf q}) }\over{E+io-{E_{\bf q}}_{\pi}-
{E_{\bf q}}_{N} }}\eqno(5.14)$$
where
$M_{\Delta}=1322MeV$ is the "bare" $\Delta$ mass

$$h({\bf q})={g\over{(1+{{{\bf q}^2}\over{\alpha^2}})^2}}\eqno(5.15)$$
and $\alpha=2.20fm^{-1};\ \ \ \ \ g=1.79{m_{\pi}}^{-3/2}$.
Thus the $\Delta$ propagator in the 
Heller-Kumano-Martinez-Moniz separable potential model 
 has the form

$$S^{\alpha\beta}_{{\bf p}_{\Delta}}(E)=
{{ u^{\alpha}({\bf p}_{\Delta}){\overline u}^{\beta}({\bf p}_{\Delta})
}\over{E-M_{\Delta}-{\tilde \Sigma}_{\pi N}(E)}}
\ {{m_{\Delta}}\over{E_{{\bf p}_{\Delta}} }}.\eqno(5.16)$$

\medskip

\begin{center}
 {\bf 6.  The results
for the $\gamma p\to\gamma' p'$,
$\gamma p\to{\pi^o}' p'$ and $\gamma p\to\gamma'{\pi^o}' p'$
observables in the $\Delta$ resonance region.}

\end{center}

\medskip

In this section we will examine the dependence of 
the observables of the 
$\gamma p\to\gamma' p'$,
$\gamma p\to{\pi^o}' p'$ and $\gamma p\to\gamma'{\pi^o}' p'$
reactions on the different propagators of the  $\Delta$ resonance. 
In the second part of this section we will
consider the sensitivity of the cross sections of the 
$\gamma p\to\gamma'{\pi^o}' p'$ reaction to the magnitude of the
$\Delta^+$ magnetic moment in different kinematical regions.

Our numerical calculations are restricted to the one particle
($N$, $\Delta$, $\pi$ and $\rho,\omega$) 
exchange model which was applied in most of modern investigations
of these reactions. Unlike to other investigations
we will calculate all three $\gamma p$ reactions with the
same input vertex functions. Besides, we will 
 take into account retardation effects and use the Coulomb gauge.
The drawback of the considered one particle exchange model
for the multichannel $\gamma p$ 
reactions is  the violation of the unitarity condition. 
Generalizations of the $\pi N$ scattering equations \cite{M1,MR}
for the coupled $(\pi N,\gamma N,\gamma \pi N)$ channels 
including unitarity has not yet been done. Such an investigation  
seems to be a necessary step for the unified and quantitative description
of the multichannel $\gamma p$ (as well as $\pi N$) scattering reactions.
Besides on this stage of our investigations we have not taken into account
the contributions of nonresonant $\pi N$ partial waves and antinucleon
degrees of freedom.
Therefore in the present paper we consider 
 only qualitative effects in the multichannel $\gamma p$ reactions. 
\newpage
 
{\bf \underline {Compton scattering on the proton }}. 
\medskip

We describe
the elastic $\gamma p$ scattering reaction in the $\Delta$ resonance
region  with the six diagrams depicted in the Figure
6. The corresponding vertex function are listed in appendix A and the
different $\Delta$ propagators are defined in equations (5.10), (5.12)
and (5.16) for the corresponding models A, B and C.
Diagrams 6a and 6b describe the  $\gamma p\to\gamma' p'$ reaction 
in the $s$ and $u$ channels with intermediate $N$ and $\Delta$ states.
Diagrams in the Figs.6c and d correspond to the one $\pi^o$ exchange
in the elastic $\gamma p$ scattering reaction.
The calculation of the proton Compton scattering 
in ref.\cite{Pascalustsa} was based on the same diagrams, 
but our calculation is not restricted to the tree
approximation.
 

\vspace{5mm}

\begin{figure}[htb]
\centerline{\epsfysize=145mm\epsfbox{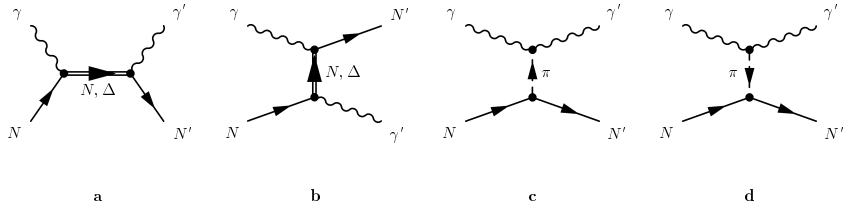}}
\vspace{-12.0cm}
\caption{\footnotesize {\it Diagrams used for the calculation of the 
Compton Scattering $\gamma p\to\gamma' p'$
 in the $\Delta$ resonance region: (${\bf a}$)
$N,\Delta$ exchange $s$-channel terms, (${\bf b}$)
$u$-channel  terms, and ${\bf c,d}$ the
$t$-channel  $\pi^o$ exchange diagrams 
with the different chronological sequence of the intermediate pion emission
and absorption. In the "old fashioned" perturbation theory the sum of
 diagrams $c$ and $d$ is equivalent to the Feynman one pi-meson exchange
 diagram, because the different time orderings
is taken into account there automatically.} }  
\label{fig:six}
\end{figure}
\vspace{5mm}

\newpage

\begin{figure}[htb]
\includegraphics[angle=-90,width=16.5cm]{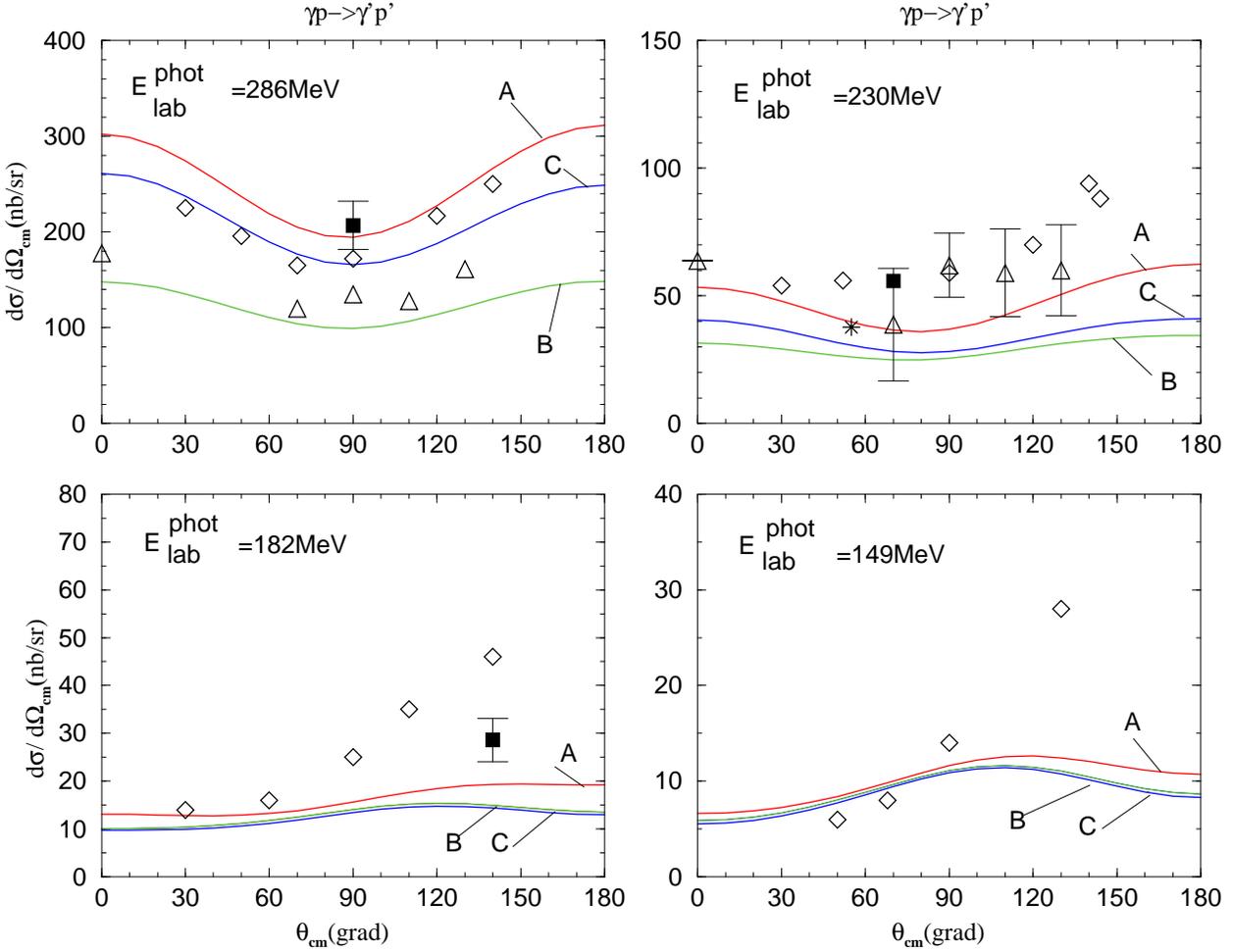}
\vspace{1.0cm}
\caption{\footnotesize {\it Variation of the differential
cross section of the proton Compton scattering
for the different
propagator of the $\Delta^+$. The curve $A$,$B$ and $C$ relate to the
expression of the $\Delta^+$ propagator (5.10),(5.12) and (5.16).
The data are from ref. \cite{Hallin} } }
\label{fig:seven}
\end{figure}
\vspace{2mm}
 
 Fig. 7 shows the differential cross section for the elastic $\gamma p$
 scattering reaction for the different energies
 of the incoming  photon ($E_{lab}^{\gamma}=149,182,230,286MeV$) and
 with the different $\Delta$ isobar propagators. The sensitivity of
 these cross sections to the form of the $\Delta$ propagators
 increases in the  $\Delta$ production region.
 Different curves on this Figure have the same qualitative behaviour.
 Therefore none of the three  $\Delta$ propagators A(5.10), B(5.12)
 or C(5.16) seems to be
  more preferable according to the present comparison with
 the experimental datas \cite{Hallin}.

\newpage

{\bf \underline {Pion photoproduction reaction}}. 
\medskip

Our calculation of the  $\gamma p\to {\pi^o}'p'$ reaction is  
based on the same set of diagrams as in ref. \cite{Blan,Dr2}, but
our calculation is performed 
in the three dimensional, time-ordering form 
in the Coulomb gauge and with retardation effects included. 
These diagrams are depicted in
 the Fig. 8. As for proton
Compton scattering, the one $N,\Delta$ exchange diagrams in 
the Fig.8a,b relates to the $s,u$ channel interaction terms.
The $t$ channel is described by $\rho$ and $\omega$ meson
exchange diagrams in the Fig.8c,d.


\vspace{0.9cm}

\begin{figure}[htb]
\centerline{\epsfysize=140mm\epsfbox{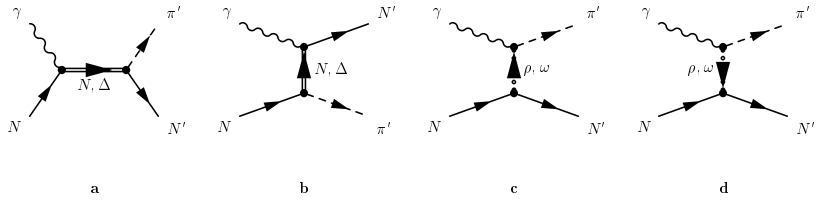}}
\vspace{-13.2cm}
\caption{\footnotesize {\it 
Pion photo-production on the proton.
One particle $N,\Delta$ and $\rho,\omega$
exchange diagrams  taken into account in the numerical calculation
of the $\gamma p-{\pi^o}' p'$ reaction.} } 
\label{fig:eight}
\end{figure}

\vspace{-1.0cm}


\begin{figure}[htb]
\includegraphics[angle=-90,width=11.5cm]{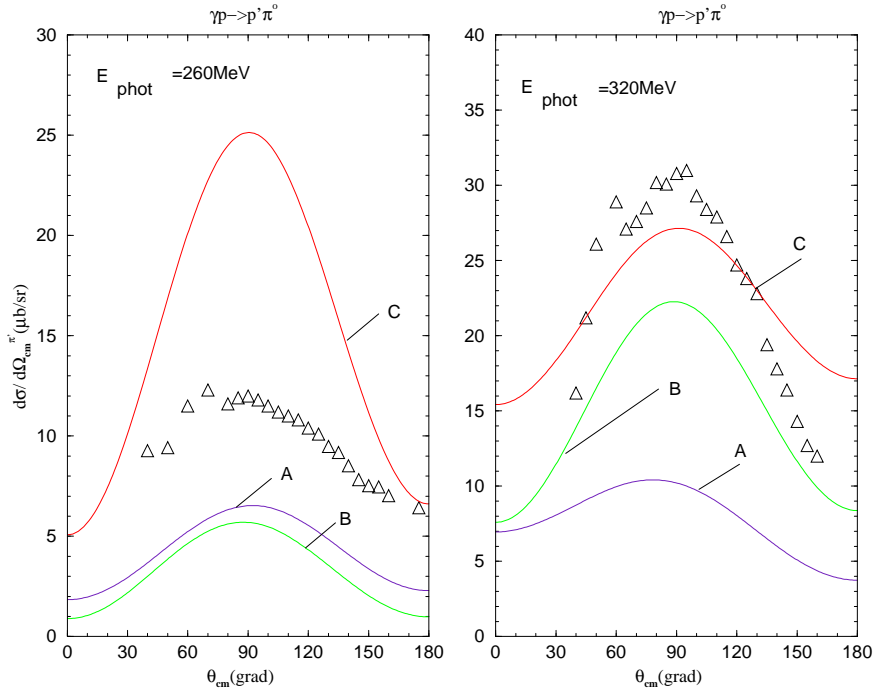}
\vspace{-0.5cm}
\caption{\footnotesize {\it 
The differential cross sections for the pion photoproduction reaction
for the $A$ (5.10), $B$ (5.12) and $C$ (5.16) $\Delta$ propagators.
The experimental results indicated by triangles are         from 
ref.\cite{Genzel}}. }  
\label{fig:nine}
\end{figure}


Fig.9 shows the dependence of the differential cross section of the
$\pi^o$ photoproduction
 $\gamma p\to {\pi^o}'p'$  on the $\Delta$ isobar propagators
A,B and C for the two energies $(E_{lab}^{\gamma}=260,320MeV)$
of the incoming photon.
As in the Compton scattering $\gamma p\to \gamma'p'$, the curves in
 the Fig.9 have qualitatively the same behavior as the experimental
observables. But unlike to proton Compton scattering, the difference
 between the different propagators
A,B,C is larger and this differences are more
 sensitive to the initial photon energy.


\medskip
{\bf \underline { The $\gamma p\to \gamma' {\pi^o}'p'$ reaction}}. 
\medskip

We now turn to the the $\gamma p\to\gamma' {\pi^o}'p'$ reaction which
we have calculated with the same vertex
 functions as the 
$\gamma p\to \gamma'p'$   and $\gamma p\to {\pi^o}'p'$ reactions
(see Appendix A). For this calculation
 we have used the diagrams depicted in the  Figure 10.

\vspace{0.7cm}

\begin{figure}[htb]
\centerline{\epsfysize=135mm\epsfbox{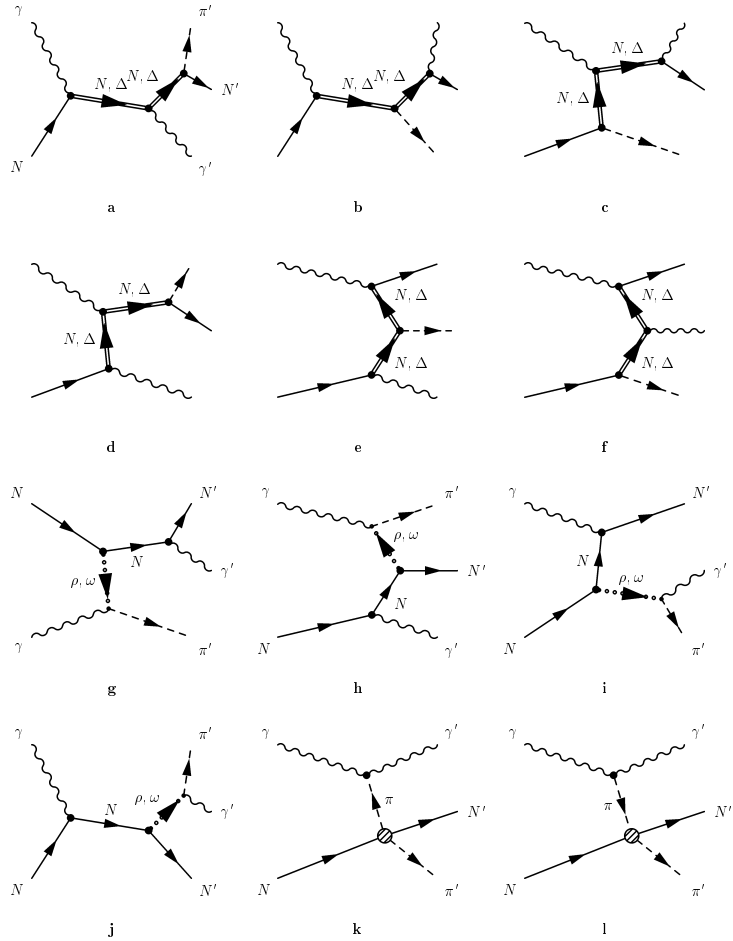}}
\vspace{-3.0cm}
\caption{\footnotesize {\it
Diagrams for the $\gamma p\to\gamma'{\pi^o}' p'$ reaction with
one particle $N,\Delta$ and $\pi^o$, $\rho,\omega$
exchange which are taken into account in our numerical calculation.
For the diagrams ${\bf b}$ and ${\bf e}$ contributions of the $\pi$-meson 
creation from the intermediate $N$ or $\Delta$ in the transitions $(N,N)$
$(N,\Delta)$ and $(\Delta,N)$ are included, but not the $\pi^o$ creation 
in the $(\Delta,\Delta)$ transition. In ref.\cite{Mink} it is shown that this 
contribution is weak.
In the $\rho,\omega$-meson exchange diagrams ${\bf g,h,i,j}$ only
 nucleon but not the $\Delta$ exchange is taken into account. In the 
one $\pi^o$-exchange
diagrams ${\bf k,l}$ the dashed circle indicates the $\pi N$ scattering 
$t$-matrix. } }  
\label{fig:ten}
\end{figure}

\newpage

For the  one-$\Delta$ 
exchange diagrams we have calculated  
diagrams in the Fig. 10a,b,c,d,e,f, i.e. we have taken into account
diagrams with $\Delta-\pi N$,
$\Delta-\gamma N$, $\Delta-\gamma \Delta$ transitions and we 
have omitted diagrams with $\Delta-\pi \Delta$ vertex.
In our calculation we have included  diagrams with
$\rho,\omega$ exchange (Fig.10g,h,i,j) which gives the important
contributions in the $\gamma p\to {\pi^o}'p'$ reaction and also 
diagrams with one $\pi^o$ exchange (Fig.10k,l) which are important for
the $\gamma p\to \gamma'p'$ reaction. 
The $\pi N$ scattering $t$-matrix in the Figs.10k,l
 is approximated
by the $N,\Delta$-exchange $s,u$-channel terms. Due to the small 
$\pi^o$ decay coupling constant (see eq. (A.9) in Appendix A),
contributions of the $\pi^o$ exchange 
diagrams ( Fig.10k,l) are small (less as $1\%$ in our calculation 
of the corresponding cross section).

The main goal of our calculation of the reaction
$\gamma p\to\gamma'{\pi^o}'p'$ is to estimate the
contributions of background diagrams which are mixed with
the diagram with the $\Delta-\Delta \gamma'$ transition
(Fig.10a). This diagram contains the interesting value of the $\Delta^+$ 
magnetic moment $\mu_{\Delta}$ (eq. (A.7) in Appendix A) and  gives the most
important  contribution for the 
determination of the magnetic moment of the $\Delta^+$
resonance. An other diagram with a $\Delta-\Delta\gamma$ transition
is depicted in Fig.10d. But the
contribution of this diagram is not important for the  cross 
sections. The complete number of the calculated diagrams is
38 (four diagrams of Fig.10a,b,c,d; $-2$ 
diagrams with
the $\Delta-\pi\Delta$ transition in Fig.10b,e;  $2\times 4=8$ diagrams
in Fig.10g,h,i,j and two diagrams in Fig.10k,l with $s,u$-channel
$N,\Delta$ exchange $\pi N$ interaction.
$4\times6-2+2\times 4+2\times 4=38$).

The first questions in estimating the background diagrams is:
what is the contribution of the diagrams with $N-\gamma'N$ transition
which generates an infrared (bremsstrahlung with a
$1/E_{\gamma'}$ energy dependence) behaviour in the cross sections? 
In order to answer this question, let us consider Fig.11, where 
the 
cross section $d\sigma/dE_{\gamma'}d\Omega^{\gamma '}_{cm}$
with 
$E_{\gamma}^{lab}=348,398,449 MeV$, 
$\theta_{cm}^{\gamma'}=110grad;\phi_{cm}^{\gamma'}=0$ 
and for the Breit-Wigner type
propagator (model B, eq. (5.12)) are shown. The full curve includes the
contributions of all diagrams in Fig.10, the long-dashed curve
corresponds
to the contribution of the single diagram with $\Delta-\gamma'\Delta$
transition (Fig.10a with intermediate $\Delta's$) and long dashed curve
includes the contributions of all diagrams without infrared $p-\gamma'p'$
transition. Thus from the Fig.11 we see that the contribution of the
term with the interesting $\Delta-\gamma'\Delta'$ transition (Fig.10a)
is comparable with 
the contributions  of all other diagrams
only at energies of the emitted $\gamma'$  at
around $80 MeV$  ($E_{\gamma'}^{cm}\simeq 80 MeV$).
The contribution of this diagram 
is further increased by increasing of the energy of the initial photon.
The contribution of the diagram $\Delta\to\Delta'\gamma'$
in Fig.11, which is proportional to the magnetic moment of the
$\Delta^+$ is most important for the
following direction of the emitted photon 
$\theta_{cm}^{\gamma'}=110grad;\ \phi_{cm}^{\gamma'}=0$.
Thus  the most preferable kinematical region for
the investigation of the role of the $\Delta-\gamma'\Delta$ transition
in the $\gamma p\to\gamma'{\pi^o}'p'$ reaction is   
$E_{\gamma'}^{cm}> 80 MeV$ and 
$\theta_{cm}^{\gamma'}\sim110grad;\ \phi_{cm}^{\gamma'}\sim0grad$.



\begin{figure}[htb]
\includegraphics[angle=-90,width=14.0cm]{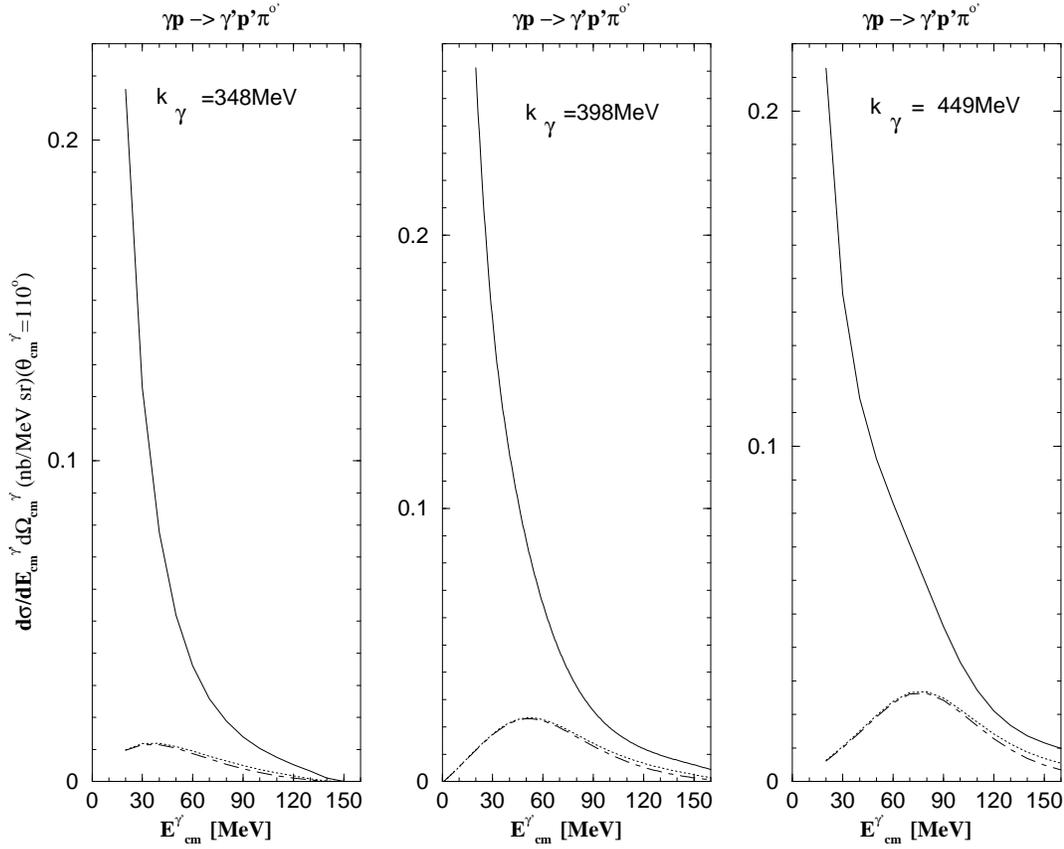}
\vspace{-0.5cm}
\caption{\footnotesize {\it
The  cross section $d\sigma/dE_{\gamma'}d\Omega^{\gamma '}_{cm}
[nb/MeVsr]$ 
of the $\gamma p\to\gamma'{\pi^o}'p'$ reaction with the 
$\Delta$ propagator of the Breit-Wigner shape (5.12)
and for different energies of the incoming photon 
$|{\bf k}_{\gamma}|\equiv E_{\gamma}$. The dashed line corresponds to 
our calculation without $p-\gamma'p'$ or $p'-\gamma'p$.
The long-dashed line is our result with only the diagram with the 
$\Delta-\gamma'\Delta'$ transition. The full curve includes the contributions
of all diagrams in Fig.10.} }
\label{fig:eleven}
\end{figure}
\vspace{5mm}

\newpage

In Fig.12 the same cross sections as in Fig.11 are displayed
but with different $\mu_{\Delta}=2.79\mu_N$ and 
$\mu_{\Delta}=2\times2.79\mu_N $ magnetic moments of the $\Delta^+$
resonance. 
The difference between corresponding curves by 
$E_{\gamma}^{lab}=348$ and $398MeV$ is quantitative and roughly
no more than $10\%$. But for the $E_{\gamma}^{lab}=449MeV$ this difference
 is  more significant.

\begin{figure}[htb]
\includegraphics[angle=-90,width=14.0cm]{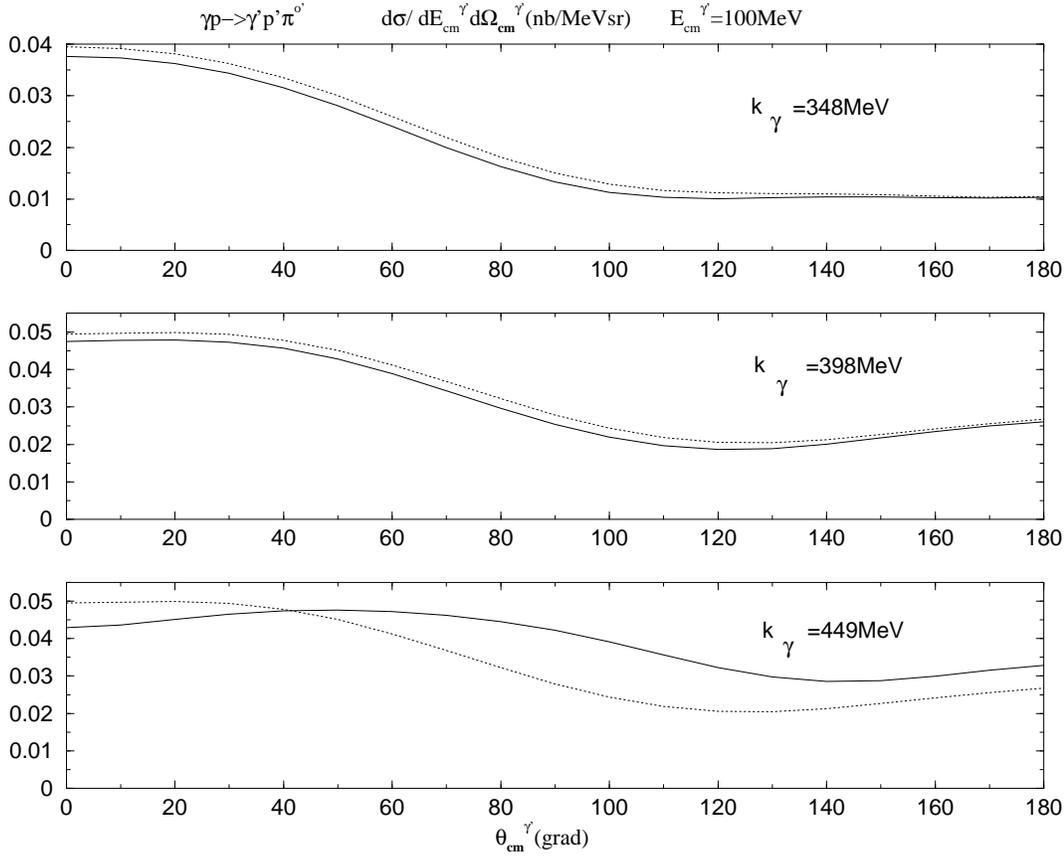}
\vspace{1.0cm}
\caption{\footnotesize {\it
The same cross section as in the Fig.11 but with a fixed energy of the emitted
photon and variations of the photon emission angle.
The dotted line by the $\theta_{[cm]}^{\gamma'}=0\ grad$  corresponds to
$\mu_{\Delta}=2.79\mu_N$
and solid line   relates to $\mu_{\Delta}=2\times 2.79\mu_N$.} }  
\label{fig:twelve}
\end{figure}

\newpage

In Fig.13 and Fig.14 we show the cross sections 
$d\sigma/d{E^{\gamma '}}_{cm}$ and  $d\sigma/d{\Omega^{\gamma'}}_{cm}$ 
for the different energies of the initial photon
($E_{\gamma}^{lab}=348$, $398$ and $449 MeV$ corresponds to 
$s^{1/2}=1239$, $1277$ and $1318MeV$) and 
with the different $\Delta$ propagators 
from the model A,B,C of section 5. The difference between
corresponding curves is large and most important for  small
. But the sensitivity of these curves 
on the different values of the $\Delta^+$ magnetic moment is small
$<10\%$. The exception is in Fig.14 for the total energy
 $s^{1/2}=1239MeV$ corresponding to $ {E^{\gamma}}_{lab}=348MeV$,
where the $\Delta$ propagator of model C (5.16) was used
for magnetic moments $\mu_{\Delta^+}=2.79$
and $2\times 2.79[nuclear\ magnetons]$.


\begin{figure}[htb]
\includegraphics[angle=-90,width=14.0cm]{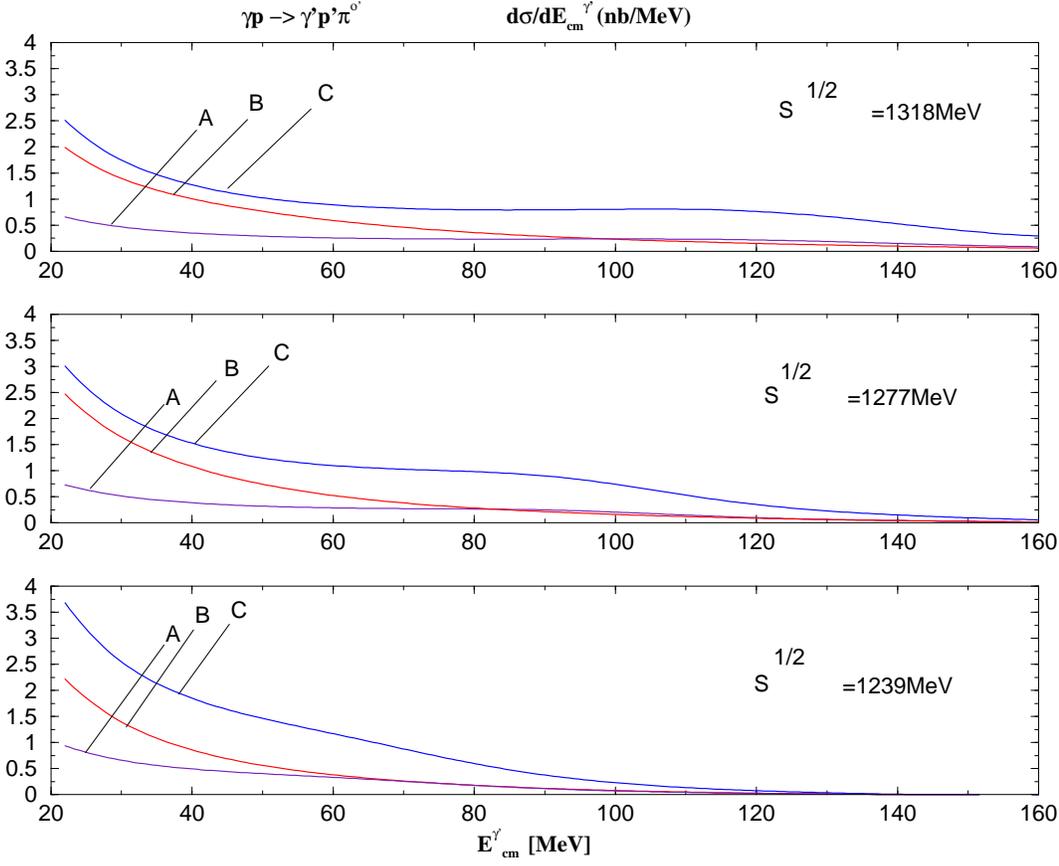}
\vspace{0.5cm}
\caption{\footnotesize {\it
Cross section $d\sigma/d{E^{\gamma '}}_{cm}[nb/MeV]$ 
as function of the 
energy of emitted  photon $\gamma'$ for the 
 incident photon energies $E_{\gamma}=348,\ 398$ and $449$MeV 
for the different $\Delta$ propagators $A$ (eq.(5.10)),
$B$ (eq.(5.12)) and $C$ (eq.(5.16)) 
For the magnetic moment 
$\mu_{\Delta^+}=2.79[nuclear\ magnetons]$ has been assumed.
}}
\label{fig:triteen}
\end{figure}
%

\newpage

The curves in Fig.13 and 14 qualitatively describe the experimental
data which has been measured recently by the A2/TAPS collaboration at
MAMI{\cite{Kotull}. But the sensitivity of the calculated cross section  
$d\sigma/d{E^{\gamma '}}_{cm}$ and  $d\sigma/d{\Omega^{\gamma'}}_{cm}$ 
on the magnitude of the $\Delta^+$ magnetic moment is even smaller
as in the corresponding calculation in ref.\cite{Dr2}.
This difference can be explained with the different gauge 
conditions, different number of included diagrams, with the missing 
retardation effects in ref.\cite{Dr2} etc. 
Only more complete calculations of the multichannel
$\gamma p$ scattering equations with unitarity
and a more consistent model of the $\Delta$ propagator, with
rescattering effects in the nonresonant $\pi N$ interactions and
antinucleon degrees of
freedom  can quantitative determine the interesting differential 
cross sections.
Keeping in mind, that the cross sections of the $\gamma
p\to\gamma'{\pi^o}'p'$
reaction are much more sensitive to the form of the $\Delta$ particle
propagator as to the magnitude of the $\Delta^+$ resonance, there
appear the next question: is the present sensitivity of these
differential cross sections enough for
determination of the magnetic moment of the $\Delta^+$ resonance?
Or in other words, exists a kinematical region where
results of our calculation are qualitatively depending on $\mu_{\Delta^+}$? 
In order to find the kinematical region, where the
the dependence on $\mu_{\Delta}$ is largest, we will consider the 
following cross sections:

\begin{figure}[htb]
\includegraphics[angle=-90,width=14.0cm]{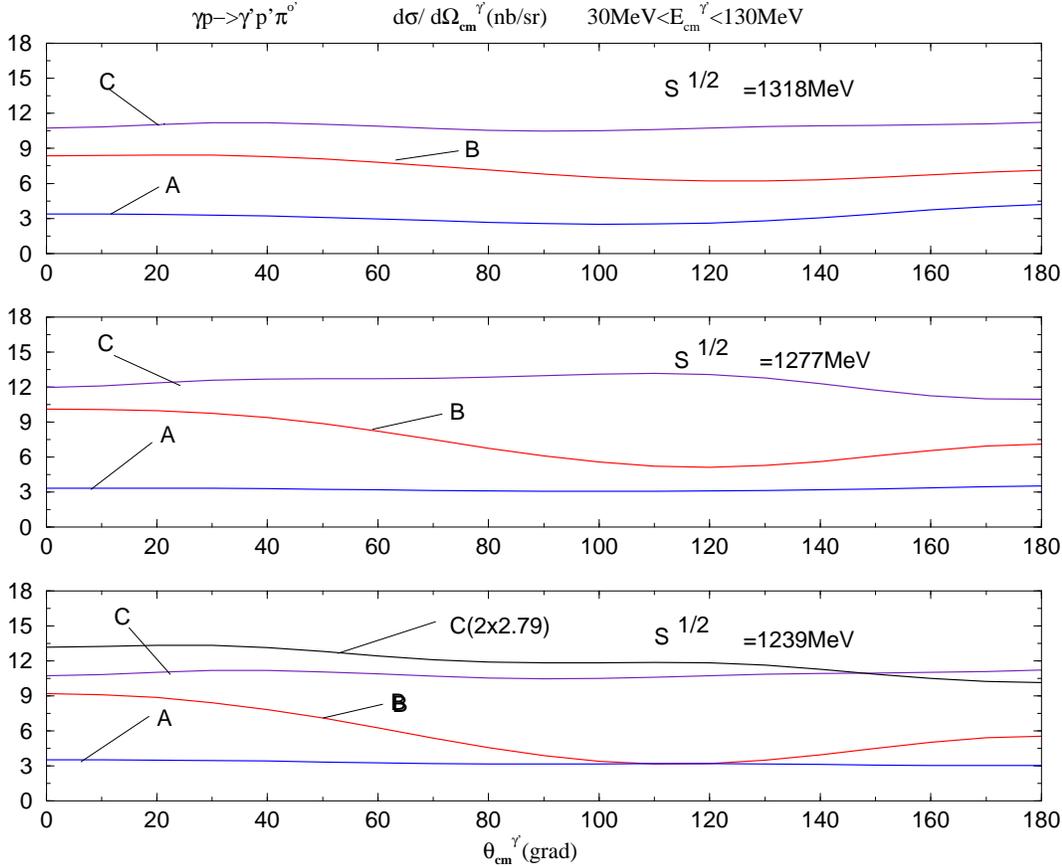}
\vspace{0.5cm}
\caption{\footnotesize {\it
Angular distribution $d\sigma/d {\Omega^{\gamma'}}_{cm}[nb/sr]$
of the final photon $\gamma'$. The different curves  
represent the three different approaches for the $\Delta$ propagator
and three different initial photon energies
as in the previous Figure.
For $s^{1/2}=1239MeV$ and $\Delta$ propagator of model $C$ (eq.(5.16))
the angular distribution of the emitted $\gamma'$ is calculated for
two magnetic moments $(2.79\ and\ 2\times 2.79)[nuclear\ magnetons]$ 
of the $\Delta^+$ resonance.} }
\label{fig:fourteen}
\end{figure}
\vspace{0.7mm}


In Fig.15 we show the sensitivity of the angular distribution 
 $d\sigma/d\Omega^{\gamma '}_{cm}$ to the A (5.10), B
(5.12) and C (5.16) models of $\Delta$ propagators and to the 
three different value
of the $\Delta^+$ magnetic moment $\mu_{\Delta^+}=0,\ 3\mu_N,\ 6\mu_N$.
Calculations are performed for two values of the initial photon
energy
${E^{\gamma}}_{lab}=400MeV$ and $450MeV$, where the energies of final
photon are integrated over the following intervals:
$80<{E^{\gamma'}}_{cm}<150MeV$ and $100<{E^{\gamma'}}_{cm}<150MeV$.
For ${E^{\gamma}}_{lab}=400MeV$ 
the variation of $\mu_{\Delta^+}$ gives an essential
difference of about
$\sim 20\%$ for the model C of the $\Delta$ propagator. 
This difference decreases for  
${E^{\gamma}}_{lab}=450MeV$. In this case the cross sections
for the $\mu_{\Delta^+}=0$ and $3\mu_N$ practically coincides. This is
 different to the case with ${E^{\gamma}}_{lab}=400MeV$, 
where the curves with $\mu_{\Delta^+}=0$ and $6\mu_N$ are close to
 each other. But one obtains a different result for $\mu_{\Delta^+}=3\mu_N$.  
The values of the cross sections A and B are close to the 
experimental data \cite{Kotull}, but these curves are less
dependent on $\mu_{\Delta^+}$. In addition the behaviour of cross
sections A are quantitatively different for the
${E^{\gamma}}_{lab}=400MeV$ and for the ${E^{\gamma}}_{lab}=450MeV$.

\vspace{0.7cm}

\begin{figure}[htb]
\includegraphics[angle=-90,width=14.0cm]{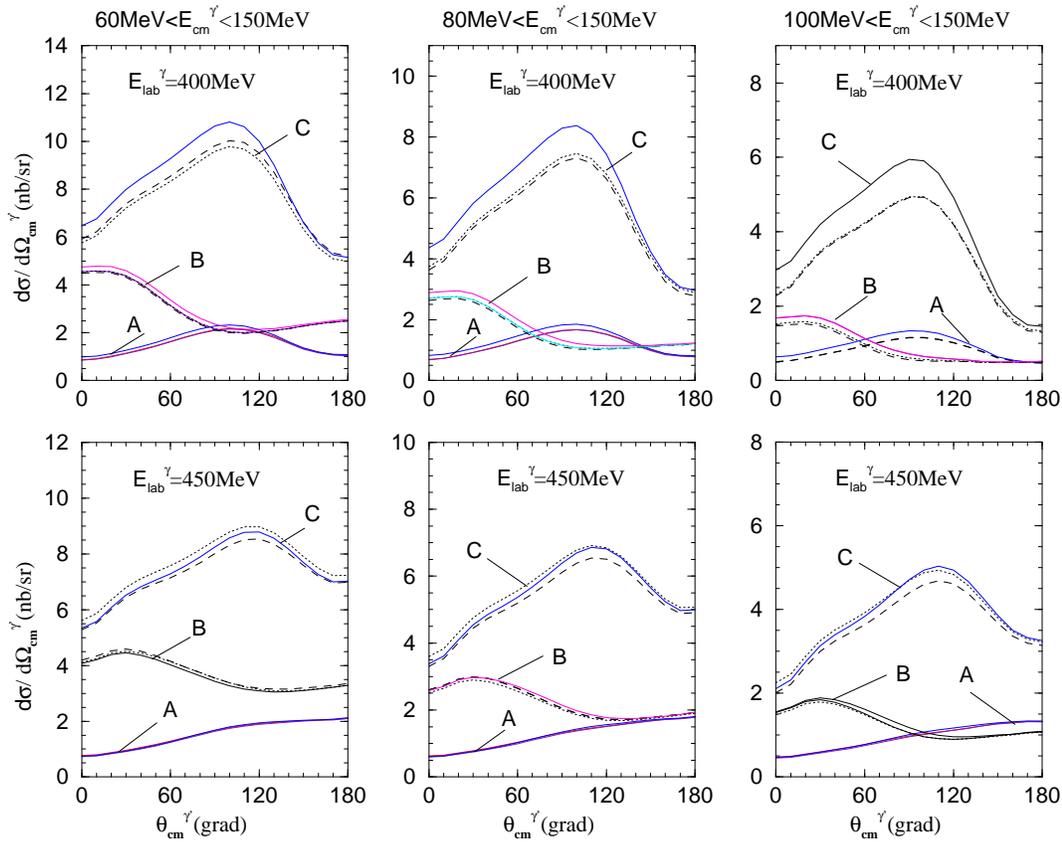}
\vspace{0.3cm}
\caption{\footnotesize {\it
Variation of the angular distribution 
$d\sigma/d\Omega^{\gamma '}_{cm}[nb/sr]$ 
of the final photon $\gamma'$energies,  for 
different  energies of the initial  photon $\gamma$, 
for different
propagators of the $\Delta$ (A,B,C see text) and for different
values of the magnetic moments of the $\Delta^+$. The dashed line corresponds 
to $\mu_{\Delta}=0$. The full curve for the values $\mu_{\Delta}=3\mu_N$
and dotted line relates to the $\mu_{\Delta}=6\mu_N$.
The energies of the final photons 
${E^{\gamma'}}_{cm}$ are integrated over
different intervals: 
$60<{E^{\gamma'}}_{cm}<150MeV$,
$80<{E^{\gamma'}}_{cm}<150MeV$ and $100<{E^{\gamma'}}_{cm}<150MeV$ for two
initial photon energies: ${E^{\gamma}}_{lab}=400MeV$ and 
${E^{\gamma}}_{lab}=450MeV$.} }  
\label{fig:fiveteen}
\end{figure}
\vspace{3mm}


\begin{figure}[htb]
\includegraphics[angle=-90,width=14.0cm]{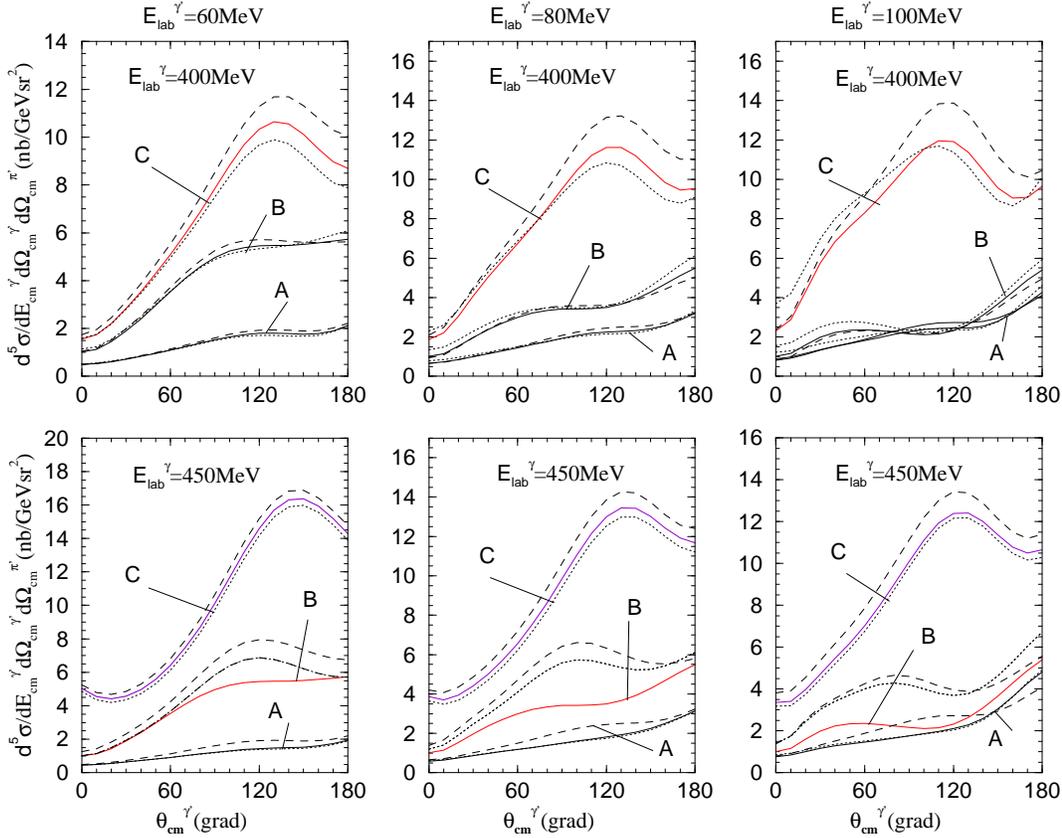}
\vspace{0.5cm}
\caption{\footnotesize {\it
The angular distribution  for the 
five-fold cross section $d^5\sigma/dE^{\gamma'}_{cm}
d\Omega^{\gamma '}_{cm}d\Omega^{\pi'}_{cm}[nb/GeVsr^2]$ 
for the following angles of the final photon and pion 
$\phi_{\gamma'}=0^o$, ${\theta^{\pi'}}_{lab}=15^o$ and $\phi_{\pi'}=0^o$. 
} } 
\label{fig:sixteen}
\end{figure}

\vspace{3mm}

The sensitivity of the cross sections to different models of 
$\Delta$ propagators and to
the $\Delta^+$ magnetic moments
$\mu_{\Delta^+}=0,3\mu_N,6\mu_N$ is examined also 
in the Figure 16, where  the five-fold cross sections 
$d^5\sigma/d^3{\bf k'}^{\gamma'}_{cm}d\Omega^{\pi'}_{cm}$ 
with fixed values of the scattering angles
$\phi_{\gamma'},{\theta^{\pi'}}_{lab},\phi_{\pi'}$ and the emitted
photon energy are shown. Unlike to the previous Figure, here the difference
between the curves with different $\mu_N$ is more visible. Most
promising is here the quantitative difference between differential
cross sections for the model B with ${E^{\gamma}}_{lab}=450MeV$
and $\mu_{\Delta^+}=0$, $\mu_{\Delta^+}=3\mu_N$ and
$\mu_{\Delta^+}=6\mu_N$.
This difference is not only very large, but also quantitatively different
as for the incident photon energy ${E^{\gamma}}_{lab}=400MeV$.

Thus we see that in special kinematical regions
the sensitivity of differential cross sections of the 
$\gamma p\to\gamma'{\pi^o}'p'$ reaction to the magnitude  
$\mu_{\Delta^+}$ can be large and this difference can have a
qualitative nature. The corresponding kinematical region is different 
for the different $\Delta$ propagators. But the region
with the ${E^{\gamma}}_{lab}>400MeV,{E^{\gamma'}}_{lab}>80MeV\
and\ {\theta^{\gamma'}}_{cm}\simeq 110^o$ is most preferable 
for the determination of the $\Delta^+$ magnetic moment $\mu_{\Delta^+}$.

\newpage

\medskip

\begin{center}
                  {\bf 7. Conclusion}
\end{center}

\medskip

The present paper is devoted to the unified field-theoretical
formulation 
of the multichannel 
$\gamma p\to\gamma' p'$, $\gamma p\to{\pi^o}' p'$ and 
$\gamma p\to\gamma'{\pi^o}' p'$ reactions and their  calculation
in the framework of the one-particle exchange model.
The field theoretical formulation was carried out in the framework
of the ``old fashioned perturbation theory'' or spectral decomposition
method over the asymptotical (Fock space) states. The present relativistic
formulation has the following attractive features:

\begin{itemize}

\item{\bf 1.}
Unlike other relativistic approaches \cite{Dr2},
our resulting amplitudes of the $\gamma p$ multichannel reactions
takes into account all retardation effects. The
 formulation is from the beginning
three dimensional and therefore it is free from the ambiguities 
which are appearing by the 
three dimensional reduction of the four dimensional Bethe-Salpeter
equations \cite{M1,M2,M3}. 
   The use of the three-dimensional relativistic equation derived in the
   framework of the old perturbation theory 
    is convenient, because in this formulation the
   scattering amplitudes have a minimal off-shellness and the off-shell
   contributions of external and internal particles can be distinguished
   in the  potential. Moreover, in the three-dimensional relativistic
   approach considered here, the effective potential
   (or exchange
   currents) are constructed from one-variable vertex functions, i.e.,
   vertex functions with two on-mass shell particles. In all other
   field-theoretical approaches the effective potentials (or exchange
   currents) are defined by vertices depending on two or three variables.
   Therefore in these formulations  severe
   approximations are necessary: there  nucleon and $\Delta$ off-mass 
   shellness is usually neglected.
\item{\bf 2.}
The complete set of the time-ordered diagrams of the 
$\gamma p\to\gamma'\pi' N'$ reactions with off mass shell 
$\gamma,\gamma'$ and $\pi'$
is presented and analyzed.
The general  structure  of the corresponding diagrams and  scattering
amplitudes does not depend
on the choice of the model of an effective Lagrangian. The
field-theoretical equations considered above are exactly connected with  
all other field-theoretical equations i. e. we can derive 
the Bethe-Salpeter equation
in the framework of the $S$-matrix reduction technique.
Therefore, all results obtained
in the framework of this time-ordered, three-dimensional  equations remain
valid in other field-theoretical approaches as well.
\item{\bf 3.}
It was shown that
 in the suggested equations for 
amplitudes of the  $\gamma N\to\gamma'{\pi^o}' N'$ reaction
 with Coulomb gauge the current conservation 
condition is automatically satisfied if the requirement of the
current conservation   for the photon-hadron vertex
functions is fulfilled. Therefore, unlike in refs. 
\cite{Blan,Koch,Ohta,Haberz,Pascalustsa,Castro,Dr2},
it is not necessary to restrict the number of 
the calculated diagrams,  or to combine some diagrams in the tree 
approximation, or to make  additional assumptions about the 
$\Delta$ propagator
in order to ensure the current conservation.
\item{\bf 4.}
The separable model of the $\pi N$ interaction
is generalized for the case of the construction of the 
spin $3/2$ particle propagator of the $\Delta$-resonance.
This procedure allows us to obtain the $\pi N\to\Delta$ form-factor
and   $\Delta$ propagator directly from the  $\pi N$ $P_{33}$ phase-shifts
and afterwards use these spin $3/2$ particle propagator in the 
 microscopic calculations.

\end{itemize}

The numerical calculations of the differential cross section of the 
$\gamma p\to\gamma' N'$, $\gamma p-\pi' N'$ and 
$\gamma p\to\gamma'{\pi^o}' p'$ reactions are performed 
in the framework of the one-particle $N$,$\Delta$ and $\pi$,$\rho$,$\omega$
exchange model with two different
separable models of the $\Delta$ propagator and with the Breit-Wigner
 propagator.  
The main numerical result is that
the description  of the multichannel $\gamma p$ scattering reactions
in the $\Delta$ resonance region 
is strongly dependent on the choice of the form of the
$\Delta$-propagator. Moreover, 
the difference between  cross sections 
of the $\gamma p\to\gamma'{\pi^o}' p'$ reaction with different 
$\Delta$ propagators in the special kinematical region
is larger than for the $\gamma p\to\gamma' p'$
and $\gamma p\to {\pi^o}' p'$ reactions.
This result makes it necessary to examine the theoretical model
of the $\Delta$ resonance  and vertex functions 
based on the $\gamma p\to\gamma'{\pi}' N'$ reactions in addition
to the photon Compton scattering and pion photoproduction reactions.

The sensitivity  of the reaction 
$\gamma p\to\gamma'{\pi^o}' p'$ to the different values of
the $\Delta^+$ magnetic moment $\mu_{\Delta^+}$ is examined. 
This sensitivity is less
than $10\%$ for most differential cross sections,
measured in ref.\cite{Kotull}. However it was demonstrated that for
every $\Delta$ propagator  some special kinematical region exists,
where differences between calculated cross sections with different
 $\mu_{\Delta^+}$ are qualitative and 
yild even an effect of more than $25\%$.
This findings make it possible to extract 
in future with more improved calculations
the magnitude of the 
$\mu_{\Delta^+}$ from the experimental data of
$\gamma p\to\gamma' N'$, $\gamma p-\pi' N'$ reaction.

\medskip

\begin{center}
                  {\bf Acknowledgment}
\end{center}

\medskip

Authors thank D. Drechsel, M.I. Krivoruchenko and M. Vanderhaeghen for
discussions. We would like to express our gratitude to M. Kotulla
and V. Metag for the current interest to this work and for useful
remarks.

\medskip

\begin{center}
                  {\bf Appendix A. Vertex functions.}
\end{center}

\medskip

In the considered formulation nucleon and $\Delta$ isobar are
defined on mass shell i.e.  $N$ and
$\Delta$ are included only in the bracket vector as ordinary 
one-particle states. 
Therefore the three particle vertex functions with nucleons
and $\Delta$ isobars
are depending only on  the  four-momentum transfer $t$. 
In our calculation we have used the following vertex functions: 

{\bf \underline {The $\gamma N-N$ vertex function  }} in the Coulomb gauge
is given in eq. (3.2) and (3.4)\cite{BD}. The exact form of the
formfactors $F_{1,2}(t)$ is considered, for example in ref. \cite{Gianini}.
In our calculation for the photon-proton vertex function
we have taken $F_1(t)=F_2(t)\equiv f(t)=(1-t/a^2)^{-2}; 
\ \ \ a=0.249fm;\ \ \ \mu_N=1.79.$

{\bf \underline {$\pi N-N$ vertex function  }} is taken from the dispersion
relation analysis  \cite{Ban}

$$<{\bf p'}_N|j_{\alpha}(0)|{\bf p}_{N}>=iG(t) 
{\overline u}({\bf p})\gamma_5\tau_{\alpha} u({\bf p});\ \ \
G(t)=g_{\pi N}\Bigl(1+{{t(t-4m_N^2)}\over{4m_N^2m_o^2}}\Bigr)^{-1}\eqno(A.1)$$
where $m_o=8.6m_{\pi};\ \ \ g_{\pi N}=12.78$.

{\bf \underline {$\gamma N-\Delta$ vertex function of 
Jones-Scadron \cite{Scadron} }}. In this treatment 
$p_{\Delta}$ is the four vector the spin $3/2$ particles with the
real mass $m_{\Delta}$ i.e. 
$p_{\Delta}=(\sqrt{m_{\Delta}^2+{\bf p}^2},{\bf p})$ and

$$<{\bf p'}_N|J^{\mu}(0)|{\bf p}_{\Delta}>\equiv
\biggl\{<{\bf p'}_N|J^{\mu}(0)\biggr\}_{\pi N\ irreducable}
|\Psi_{{\bf p}_{\Delta}}>=$$
$${\overline u}({\bf p'}_N)\biggl[F_M(t)
\Gamma_M^{\mu\nu}(t)+F_E(t)\Gamma_E^{\mu\nu}(t)+F_C(t)\Gamma_C^{\mu\nu}(t)
\biggr] u_{\nu}({\bf p}_{\Delta})\eqno(A.2)$$
where $t=({p'}_N-p_{\Delta})^2$ and 
magnetic
$\Gamma_M^{\mu\beta}(t)$, electric $\Gamma_E^{\mu\beta}(t)$ and charged
$\Gamma_C^{\mu\beta}(t)$ Lorentz-invariant combination are defined as

$$\Gamma_M^{\mu\nu}(t)=-{3{(m_N+m_{\Delta})}
\over{2m_N \Bigl((m_N+m_{\Delta})^2-t\Bigr)}}
\epsilon^{\mu\nu\alpha\sigma}
(p'_N+p_{\Delta})_{\alpha}(p_{\Delta}-p'_N)_{\sigma}
\eqno(A.3a)$$

$$\Gamma_E^{\mu\nu}(t)=-\Gamma_M^{\mu\nu}(t)$$
$$-\gamma_5\ {3i{(m_N+m_{\Delta})}
\over{m_N \Bigl((m_N+m_{\Delta})^2-t\Bigr)
\Bigl((m_N-m_{\Delta})^2-t\Bigr)}}
\epsilon^{\mu\lambda\alpha\beta}
(p'_N+p_{\Delta})_{\alpha}(p_{\Delta}-p'_N)_{\beta}
\epsilon^{\nu\gamma\delta}_{\lambda}
{p_{\Delta}}_{\gamma}(p_{\Delta}-p'_N)_{\delta}
\eqno(A.3b)$$

$$\Gamma_C^{\mu\nu}(t)=
-\gamma_5\ {3i{(m_N+m_{\Delta})(p_{\Delta}-p'_N)^{\mu}
\Bigl[t(p'_N+p_{\Delta})^{\nu}-(p_{\Delta}-p'_N)^{\nu}
(m_{\Delta}^2-m_N^2)\Bigr]}
\over{2m_N \Bigl((m_N+m_{\Delta})^2-t\Bigr)
\Bigl((m_N-m_{\Delta})^2-t\Bigr)}}.
\eqno(A.3c)$$

Charge formfactor $\Gamma_C^{\mu\nu}(t)$
contributes in the our calculation with
the retardation. But  
 in the tree approximation, where the four vector
$q^{\mu}=(p_{\Delta}-p'_N)^{\mu}$ is replaced with  the real photon four
momentum $q^2=0$, contributions of $\Gamma_C^{\mu\nu}(t)$ disappears.

For the electric, magnetic and 
charge formfactors we take the same cut-off function  $f(t)$ as
for the $\gamma$-proton vertex function 
$$F_M(t)=F_M(0)f(t);\ \ \ F_E(t)=F_E(0)f(t);\ \ \  
F_C(t)=F_C(0) f(t)\eqno(A.4)$$
where $F_M(0)=3.2$ \cite{Dr3}, $F_E(0)=0.025F_M(0)$ \cite{Beck}
 $F_E(0)=(m_{\Delta}-m_N)/2m_{\Delta}F_C(0)$ \cite{Scadron}.
For our numerical calculation most sufficient is the
magnetic part of (A.2). The recent overview of the unified 
$N^*-N\gamma$ vertex functions is given in \cite{kriv2}.

{\bf \underline {The $\pi N-\Delta$ vertex function  }} are defined in
section 5. Thus  
in the model A {$\pi N-\Delta$ vertex function is $g({\bf p})$ (5.11),
in model B it coincides with the coupling constant (5.13) and 
 in model C $\pi N-\Delta$ vertex function is $h({\bf p})$ (5.15).

{\bf \underline {The $\gamma \Delta'-\Delta$ vertex functions}} are 
the same as
 in our previous paper \cite{M5}.
 In this case
 $Q=P'_{\Delta}-P_{\Delta}$ and $R=P'_{\Delta}+P_{\Delta}$
and $\gamma \Delta'-\Delta$ vertex function is

$$<{\bf P'}_{\Delta}|J_{\mu}(0)|{\bf P}_{\Delta}>=
{\overline u}^{\sigma}({\bf P'}_{\Delta})
V_{\sigma\mu\rho}({\bf P'}_{\Delta},{\bf P}_{\Delta})
u^{\rho}({\bf P}_{\Delta})\eqno(A.5)$$

where
$$V_{\sigma\mu\rho}({\bf P'}_{\Delta},{\bf P}_{\Delta})=
g_{\rho\sigma}\Bigl[
F_1(Q^2)\gamma_{\mu}+{{F_2(Q^2)}\over{2M_{\Delta} }}R_{\mu}\Bigr]
+Q_{\sigma}Q_{\rho}\Bigl[{{F_3(Q^2)}\over{M_{\Delta}^2}}\gamma_{\mu}+
{{F_4(Q^2)}\over{2 M_{\Delta}^3}}R_{\mu}\Bigr].\eqno(A.6)$$
 
The form factors $F_i(Q^2)$ are simply connected with the charge monopole
$G_{C0}(Q^2)$, the magnetic dipole $G_{M1}(Q^2)$, the electric quadrupole
$G_{E2}(Q^2)$ and the magnetic octupole
$G_{M3}(Q^2)$ form factors of the $\Delta^+$ resonance.
In the low energy region we can neglect the terms $\sim Q^2/4M_{\Delta}^2$,
and we keep only terms $\sim 1/M_{\Delta}$. Then the previous formula
can be rewritten in a similar form as the $\gamma$-proton vertex
function:

$$V_{\sigma\mu\rho}({\bf P'}_{\Delta},{\bf P}_{\Delta})=
g_{\rho\sigma}
G_{C0}(Q^2) {{R_{\mu}}\over{2M_{\Delta}}}+ig_{\rho\sigma}
{{G_{M1}(Q^2)}\over{2M_{\Delta} }}\sigma_{\mu\beta}Q^{\beta}\eqno(A.7)$$

In our case of soft photon emission we have approximated the form factors
in (A.7) with their pseudo-threshold values  
$G_{C0}(Q^2)\to G_{C0}(t_{ptr}=1$ and
$G_{M1}(Q^2)\to G_{M1}(t_{ptr})=\mu_{\Delta^+}$, where 
$t_{ptr}=(m_{\Delta}-m_N)^2$, $\mu_{\Delta^+}$ denotes the magnetic 
moment of the $\Delta^+$ resonance
and it is simply connected with the 
$k_{\Delta^+}$  anomalous magnetic moment of $\Delta^+$ 
$\mu_{\Delta^+}=(1+k_{\Delta^+})/2m_{\Delta}$.

{{\bf \underline {The $V\equiv\rho, \omega$ meson-nucleon vertex
functions}} in eq. (2.11) have the form

$$<{\bf p'}_N|j^V_{\mu}(0)|{\bf p}_N>={\overline u}({\bf p'}_N)
\Biggl(\gamma_{\mu}F^V_1(t)+
i{ {k_V}\over {2m_N}}\sigma_{\mu\nu}(p_N'-p_N)^{\nu}F^V_2(t)\Biggr) 
u({\bf p}_N)\eqno(A.8)$$
where $k_{\omega}=0;\ \ k_{\rho}=3.7$. Form factors $F^V_1(t)$
are replaced with their threshold values 
$F^V_1(t)\Longrightarrow g_{VNN}$ and 
$g_{\omega NN}=3\times g_{\rho NN}=15$ \cite{Guidal}.
And for the $\rho(\omega)$ decay constant we have taken the value
$g_{\omega\gamma\pi}=3\times g_{\rho\gamma\pi}=0.374$ \cite{Blan,Kriv}.

{\bf \underline {The $\pi^o$ decay  vertex
function}} in the Figure 10k,l is taken in the standard form
on the tree approximation \cite{Cheng,Pascalustsa}

$$\Gamma_{\gamma'-\pi^o\gamma}^{\mu\nu}\sim \int d^4x <0|
T\Bigl(j^{\mu}(x)j^{\nu } (0)\Bigr)|{\bf p_{\pi}}
={\bf k}_{\gamma}-{\bf k}_{\gamma'}>e^{ik_{\gamma'}x}
\approx -i{0.035\over {m_{\pi}}}\epsilon^{\mu\nu\alpha\beta}
{k_{\gamma}}_{\alpha}{k_{\gamma'}}_{\beta}.\eqno(A.9)$$

\end{document}